\documentclass{aastex6}

\usepackage{natbib}
\usepackage{xcolor}
\newcommand{\eps}[1]{\mbox{log~$\epsilon$(#1)}} 
\newcommand\species[2]{#1 {\sc #2}}
\newcommand\iso[2]{$^{\rm #1}$#2}

\def\ie{\mbox{i.e.}}
\def\eg{\mbox{e.g.}}

\def\teff{\mbox{T$_{\rm eff}$}}
\def\logg{\mbox{log~{\it g}}}
\def\vmicro{\mbox{$\xi_{\rm t}$}}
\def\kmsec{\mbox{km~s$^{\rm -1}$}}
\def\carbiso{\mbox{$^{12}$C/$^{13}$C}}

\shorttitle{Field Red Horizontal-Branch Stars}
\shortauthors{Af{\c s}ar et al.}

\begin{document}

\title{A SPECTROSCOPIC SURVEY OF FIELD RED HORIZONTAL-BRANCH STARS}

\author{Melike Af{\c s}ar\altaffilmark{1,2},
        Zeynep Bozkurt\altaffilmark{1},
        Gamze B{\"o}cek Topcu\altaffilmark{1},
        Dana~I. Casetti-Dinescu\altaffilmark{3,4,5},
        Christopher Sneden\altaffilmark{2}, and
        Gizem {\c S}ehito{\= g}lu\altaffilmark{1}
}

\altaffiltext{1}{Department of Astronomy and Space Sciences,                 
                 Ege University, 35100 Bornova, {\. I}zmir, Turkey;          
                 melike.afsar@ege.edu.tr, \\
                 zeynep.bozkurt@ege.edu.tr,
                 gamzebocek@gmail.com}
\altaffiltext{2}{Department of Astronomy and McDonald Observatory,           
                 The University of Texas, Austin, TX 78712, USA; 
                 afsar@astro.as.utexas.edu, chris@verdi.as.utexas.edu}
\altaffiltext{3}{Department of Physics, Southern Connecticut State 
                University, 501 Crescent Street, New Haven, CT 06515, USA; 
                danacasetti@gmail.com}
\altaffiltext{4}{Astronomical Institute of the Romanian Academy, str. 
                Cutitul de Argint 5, Bucharest, Romania}
\altaffiltext{5}{Radiology and Biomedical Imaging, Yale School of Medicine, 
                300 Cedar Street, New Haven, CT 06519}

\begin{abstract}

A metallicity, chemical composition, and kinematic survey 
has been conducted for a sample of 340 candidate field red horizontal 
branch stars.  
High resolution, high signal-to-noise spectra were gathered with the 
McDonald Observatory 2.7m Tull and the Hobby-Eberly Telescope echelle 
spectrographs, and were used to determine effective temperatures, surface 
gravities, microturbulent velocities, [Fe/H] metallicities, and abundance 
ratios [X/Fe] for seven $\alpha$ and Fe-group species.  
The derived temperatures and gravities confirm that at least half 
of the candidates are true RHB stars, with (average) parameters 
\teff~$\sim$~5000~K, \logg~$\sim$~2.5. 
From the $\alpha$ abundances alone the thin and thick Galactic 
populations are apparent in our sample.   
Space motions for 90\% of the program stars were computed from Hipparcos 
and Gaia parallaxes and proper motions.
Correlations between chemical compositions and Galactic kinematics clearly
indicate the existence of both thin and thick disk RHB stars.

\end{abstract}


\section{INTRODUCTION}\label{intro}

Red horizontal branch stars (RHBs) are core-helium-burning stars that 
have evolved beyond the red giant branch (RGB). 
They reside on the horizontal branch (HB) between red clump stars (RCs) $-$ 
the redder counterparts of the RHB stars $-$ and the instability strip. 
The location of stars on the HB is mainly affected by the dependency of 
H-burning efficiency on the envelope mass \citep{salaris06}: the higher the
envelope mass the redder the location of the star on the HB. 
Other parameters such as He abundance, metallicity, age, initial stellar 
mass and rotation also play important roles on the HB star locations. 
Although RHBs can be easily identified in color-magnitude diagrams (CMD) of
globular clusters (GC), it is more challenging to distinguish them in 
the field since they can be easily confused with the subgiant (SG) phases 
of high-mass former main-sequence (MS) stars. 
Field RHBs are relatively rare and have not been paid 
enough attention in the literature. 
In spite of the identification challenges,
field RHB stars are very important tools in many aspects. 
They can be used for distance determinations as standard candles, similar 
to RCs (e.g. \citealt{stanek98, udalski00, alves02, laney12, hawkins17}, and references therein).
Their kinematic and abundance distributions throughout the Galaxy 
can provide valuable information on the chemical evolution of our Galaxy
(e.g. \citealt{chen10, chen11, shi12}). 
Most importantly, they are the dig sites where light element 
nucleosynthesis and mixing products lead to surface abundance changes.
Several studies including \cite{rose85} and \cite{gilmore83} investigated 
the kinematics of field RHBs as a group and concluded that they are 
members of the thick disk component of the Galaxy. 

Chemical compositions for some of the thick disk RHBs have been
studied in detail by, e.g., \cite{tautvaisiene96, tautvaisiene97} and 
\cite{tautvaisiene01}.
In \cite{afsar12} (hereafter ASF12) we reported the first systematic stellar
parameter and chemical composition study of field stars with temperatures and 
luminosities consistent with the location of the HR diagram red horizontal 
branch but without stellar population assignment.
After analysis of spectroscopic data for about 80 stars with 
weakly-constrained color-magnitude diagram (CMD) positions near the RHB 
domain, ASF12 suggested that 18 were probable true RHBs, which they defined as 
those with effective temperatures 4900~K~$<$~\teff~$<$5500~K, surface gravities 
2.2~$\lesssim$~logg~$\lesssim$~2.8, 
and carbon isotopic ratios \carbiso~$\leq$~30.
Interestingly, the kinematics of these 18 stars suggested that 13 of them belong
to the thin disk, while only five of them have space motions 
consistent with the Galactic thick disk membership.
The metallicities of the thin disk RHBs were found to be essentially solar,
while the thick disk stars are somewhat metal-poor.
Thus ASF12 suggested that
a low velocity, high metallicity ``thin disk'' RHB population can
sometimes be found in the same RHB 
domain normally populated by thick disk, mildly metal-poor stars.

With an RHB sample of only 18 stars ASF12 could not draw general conclusions
about the relative prevalence of thin- and thick-disk members.
Therefore we have gathered high-resolution spectra of more than 300 additional
RHB candidates, and here we explore the metallicities, some abundance ratios, 
and space velocities of this more robust statistical sample.
The elements included in this paper are limited to those that can be easily
analyzed from equivalent width measurements (Si, Ca, Ti, Cr, Fe, and Ni).
In \S\ref{sample} we describe selection of the new targets.
Details of observations and spectrum reductions appear in \S\ref{obsred}.
We assemble stellar parallax information in \S\ref{parmag}, leading
to a color-magnitude diagram for our program stars, and in \S\ref{motions} we 
compute space velocities and estimate Galactic orbital parameters.
Spectroscopic atmosphere parameters are described in \S\ref{modelcomp}, and
derived abundances in \S\ref{abunds}. 
In \S\ref{corrkinchem} and \S\ref{orbitchem} we investigate relationships 
between the chemical composition and kinematic information in our RHB sample. 
Our results are discussed in \S\ref{discuss}.
A second paper \citep{afsar18} will consider abundances 
of the CNO group and other elements that require iterative synthetic 
spectrum computations, and will discuss in detail the evolutionary status
of the program stars.

\section{THE STELLAR SAMPLE}\label{sample}

In Table~\ref{tab-basic} we give basic data for our RHB candidate stars: 
coordinates, broad-band magnitudes, interstellar medium (ISM) extinction and 
reddening estimates, spectral types, and the telescope/instrument
combination that was used for each star.
Simplifying the RHB candidate names as much as possible, we have 
preferentially adopted the Hipparcos (HIP) catalog number 
if available, or
the Bonner Durchmusterung (BD) number. 
Some fainter stars have neither of these designations, so we mostly used
their Tycho Catalog (TYC) or Two Micron All-Sky Survey (\citealt{cutri03}, 
2MASS J, abbreviated here to simply J) numbers.
Finally, for a handful of stars, we adopted their single SIMBAD
\citep{wenger00} designations from either the HST Guide Star Catalog (GSC) 
or the \cite{weistrop83} (Weis) photometric survey.

To compile a complete list of RHB candidates, we mainly adopted the following 
criteria by making use of the results from ASF12 and \cite{kaempf05} (and 
references therein):
\begin{itemize}
\item 1.5 $\lesssim$ $V-K$ $\lesssim$ 2.2 
          (when a K magnitude was available)
\item 0.5 $\lesssim$ $B-V$ $\lesssim$ 1.0
\item $-$0.5 $\lesssim M_{\rm V}$~$\lesssim$ 1.5 
         (when a parallax was available)
\item \logg\ $<$ 3.5 
          (when a parallax was available)
\end{itemize}
To assemble a list of RHB candidates based on these criteria, we first started  
with the Hipparcos catalog. 
Since observations were obtained with two different 
telescopes and their echelle spectrographs at McDonald Observatory, the 
9.2m Hobby-Eberly Telescope (HET) and the 2.7m Harlan J. Smith Telescope (HJS)
(see next section), we had to restrict the apparent magnitudes ($V$) 
considering the brightness limitations of these setups:
11 $<$ $V$ $<$ 14 for 
the 9.2m HET and $V$ $<$ 11 for the 2.7m HJS.
Following ASF12 we made use of $B-V$ and $V-K$ colors during the selection of 
our targets. For the present study
we focused more on the $V-K$ color because it is
less dependent on metallicity and gravity than is $B-V$, and
therefore gives more reliable initial temperature estimates.
RHBs mainly reside in the $V-K$ range of $\thickapprox$~1.5$-$2.2 (ASF12). 
$B-V$ colors were taken into account only for the candidates without $K$ 
magnitude information. 

Besides broad-band $B$, $V$ and $K$ magnitudes, we also used Hipparcos 
parallaxes, literature spectral types, and Galactic coordinates to find 
likely RHB field star candidates for spectroscopic observations. 
Parallaxes were used to determine the preliminary absolute 
magnitudes to find out whether the absolute magnitudes of stars reside in
the adopted selection criteria. Spectral types were limited to the loci of the 
RHB stars in the H-R diagram (as described in \S\ref{intro}, also see ASF12). 
We also selected candidates with high Galactic latitudes in cases where 
lack of parallax information.
We considered a wide Galactic coordinate range to achieve a 
more complete sample without bias on stellar population.

We were able to calculate the absolute magnitudes ($M_{\rm V}$) for stars with 
parallaxes (no extinction information was included in assembling the 
preliminary RHB target list).
The absolute magnitude range for RHB stars is 
approximately $-$0.5 $\lesssim M_{\rm V}$~$\lesssim$ 1.5
(e.g. \citealt{kaempf05}). 
This restriction helped us to eliminate many stars
with magnitudes suggestive of main sequence membership 
($M_{\rm V}$~$\gtrsim$~4.0). We also used a method suggested by 
\cite{berdyugina94} in order to estimate preliminary surface gravities of 
our candidates. 
Their Eq.~1 relates gravity (\logg) to $M_V$, 
metallicity [Fe/H]\footnote{
We adopt the standard spectroscopic notation \citep{wallerstein59} that for
elements A and B,
[A/B] $\equiv$ log$_{\rm 10}$(N$_{\rm A}$/N$_{\rm B}$)$_{\star}$ $-$
log$_{\rm 10}$(N$_{\rm A}$/N$_{\rm B}$)$_{\odot}$.
We use the definition
\eps{A} $\equiv$ log$_{\rm 10}$(N$_{\rm A}$/N$_{\rm H}$) + 12.0, and
equate metallicity with the stellar [Fe/H] value.}
and effective temperature (\teff) 
by comparing the observed and predicted equivalent widths of MgH lines in 
subgiants and giants. 
The equation is independent of stellar mass and provides good
approximations to \logg\ 
(within 0.35 dex in case of our sample). 
For the candidates without parallaxes and $V-K$ colors, we relied on the 
$B-V$ colors and 
Galactic latitude information to provide rough \logg\ estimates.
45 stars that were identified as evolved stars from ASF12 are also 
included in
the present sample (see \S\ref{modelcomp}). 
Six of our program stars were observed twice
(see next section). 
About 370 stars were observed during this project and 340
of them had sufficient data quality to be analysed.

\section{OBSERVATIONS, REDUCTIONS, SPECTRUM MEASUREMENTS}\label{obsred}

From 2011 to 2014 we obtained high resolution, high signal-to-noise ($S/N$) 
spectra of our program stars with echelle spectrographs of two telescopes 
located at McDonald Observatory.

For the majority of targets we employed the 2.7m HJS and its Robert G. Tull 
Cross-Dispersed Echelle spectrograph \citep{tull95}.
Data were acquired in several observing runs dedicated to this program.
The instrumental setup was identical to that used by ASF12.
Briefly, the combination of echelle grating, camera, and a 1\farcs2 
width entrance slit yielded a spectral resolving power of
R~$\equiv$~$\lambda/\Delta\lambda$~$\thickapprox$~60,000.      
The total wavelength coverage was $\lambda\lambda$ 3400-10900, 
encompassing 63 echelle spectral orders.                                  
These orders overlap in wavelength from the blue wavelength limit until 
$\lambda$~$\simeq$~5900~\AA, and small spectral gaps exist for the
redder spectral orders.

For some targets we employed the 9.2m HET and its
High Resolution Spectrograph (HRS; \citealt{tull98}).
These data were obtained in queue observing mode by the HET staff in 
individual time slots that depended on the larger HET scheduling matrix,
rather than on specifically scheduled nights.
The instrumental setup was identical to that used by \cite{bocek16}.
There are two CCD detectors in this instrument to achieve large
spectral coverage.
The wavelength ranges were 5100$-$6900~\AA\ on the ``blue'' detector and
7000$-$8800~\AA\ on the ``red'' detector.
The spectral resolving power was $R$~$\simeq$~60,000.

The $S/N$ (per resolution element) values 
calculated near 6900~\AA\ were taken from the night reports of HET 
observations (please also see \citealt{bocek16}). The $S/N$ values 
of 2.7m HJS spectra were calculated in a similar manner as described 
in \cite{bocek15}. Overall, the $S/N$ values in the extracted spectra 
were high, averaging $\sim$170 for the whole sample, often $\gtrsim$~300, 
and rarely $<$~100. 

Out of our total stellar sample of 340 target stars, six were observed on 
two different occasions: HIP~4197, HIP~62325, HIP~71837, and HIP~72631 
(twice with the 2.7m telescope), and HIP~54858 and HIP~74058 (once each with 
the 2.7m and HET).
These duplicate observations are considered as separate
target stars in order to assess the repeatability of our analyses.

We reduced the spectra with IRAF's\footnote{
IRAF is distributed by the National Optical Astronomy Observatory, which is 
operated by the Association of Universities for Research in Astronomy (AURA) 
under a cooperative agreement with the National Science Foundation.}
$ccdproc$ task: bias subtraction, frame trimming, flat-fielding, and 
scattered light subtraction.
The HET/HRS ``blue'' and ``red'' frames were treated as individual 
observations at all reduction stages.
To remove most single-exposure radiation event anomalies, we combined all 
available integrations for each star into a single reduced frame. 
IRAF $echelle$ package tasks were used for the remaining data reduction steps,
beginning with spectral order extraction from the reduced data frames.
Th-Ar comparison lamp exposures were used for wavelength calibration. 
With the IRAF $telluric$ task, spectra of hot rapidly rotating stars 
obtained along side of the target star spectra were used to remove telluric 
O$_2$ and H$_2$O line contamination.

We followed \cite{bocek15,bocek16} in measuring the radial velocity ($RV$) 
shifts of our targets with the cross-correlation technique in IRAF's 
$fxcor$ task \citep{fitzpatrick93}.
This method compares the spectrum of interest with a template. 
We elected to create an artificial spectrum with model atmospheric parameters
of a typical warm giant (\teff~=~4900~K, \logg~=~2.5, [Fe/H]~=~0.0, 
\vmicro~=~1.2~\kmsec).
The synthetic spectrum wavelength range was 5020$-$5990~\AA.
The resulting observed $RV$s were transformed into heliocentric values
with IRAF's $rvcorrect$ task.
They are listed in Table~\ref{tab-motions}.

\begin{figure}
\epsscale{1.00}
\plotone{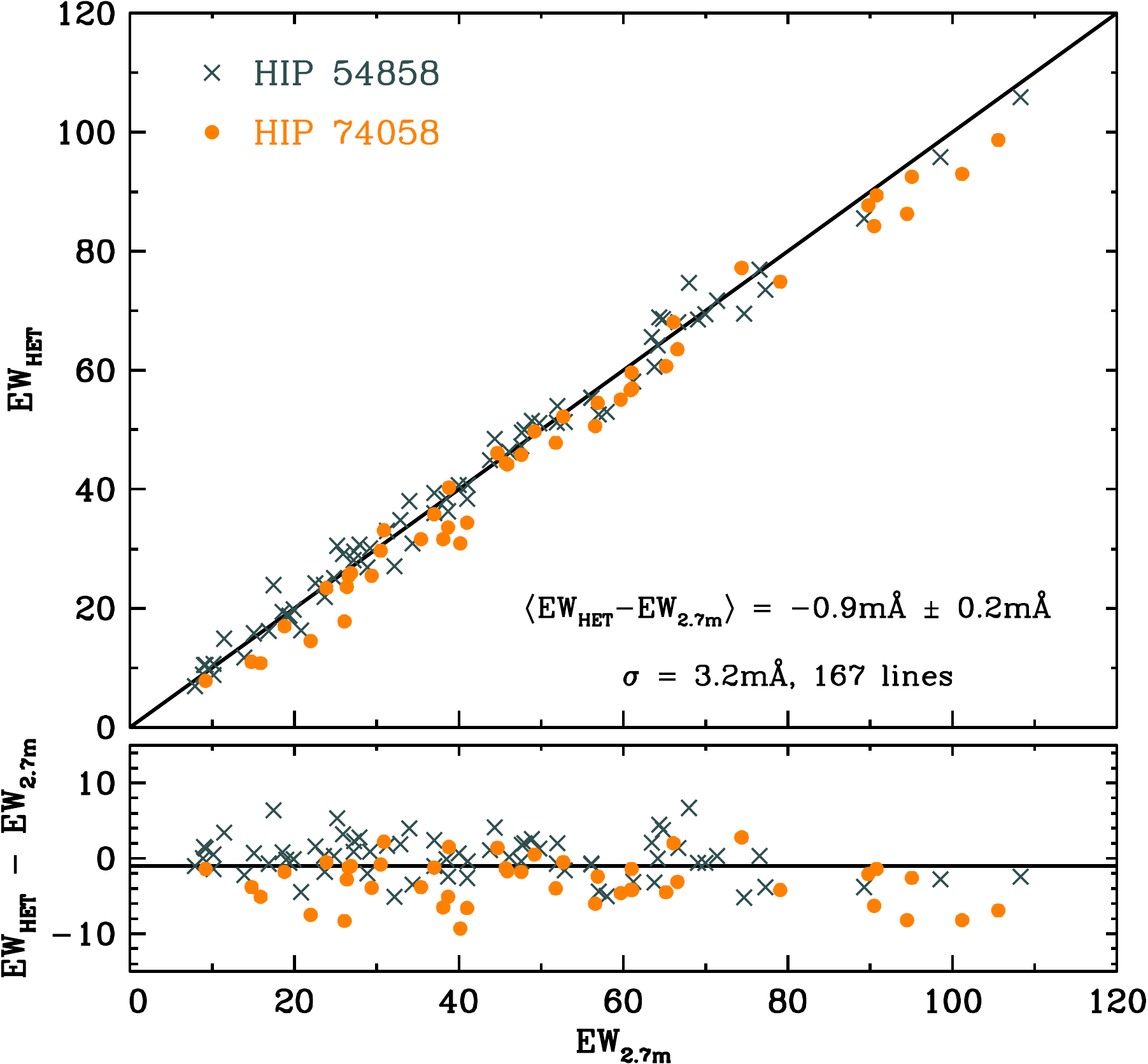}
\begin{center}
\end{center}
\caption{\label{ewcomp}
\footnotesize
    Comparison of equivalent widths for HIP~54858 and HIP~74058 measured 
    on spectra obtained with the 2.7m telescope ($EW_{\rm 2.7}$) and the HET
    ($EW_{\rm HET}$).  
    In each panel the black line denotes equality of the $EW$ measurements.
}
\end{figure}

We derived spectroscopic values of atmospheric parameters (\teff,
\logg, \vmicro, [Fe/H]), and abundance ratios [X/Fe] for many neutral
and ionized species through analyses of their equivalent widths ($EW$s).
The lines chosen for study will be discussed in \S \ref{modelcomp}.
To measure $EW$s we used an Interactive Data Language (IDL) routine 
developed by \cite{roederer10} and \cite{brugamyer11}. 
This code interactively matches observed line profiles with computed 
Gaussian line profiles in most cases, or Voigt profiles for some strong 
lines, recording both $EW$s (for model parameter and abundance determinations)
and central line depths (for initial temperature estimates).
In ASF12 we showed that $EW$s determined from 2.7m spectra for two
RHB stars were consistent with their $EW$s reported in the literature.
In Figure~\ref{ewcomp} we compare $EW$s for two program RHB stars, 
HIP~54858 and HIP~74058, measured on our 2.7m and HET spectra.
They are in accord with less than 2~m\AA\ average difference between the
two data sets.

\section{STELLAR PARALLAXES AND ABSOLUTE MAGNITUDES}\label{parmag}

Parallax data for our program stars were taken from those obtained by
the astrometric space missions Hipparcos \citep{vanleeuwen07}\footnote{
https://www.cosmos.esa.int/web/hipparcos}
and Gaia (Gaia DR1 data archive\footnote{
https://gea.esac.esa.int/archive/}, \citealt{GAIA16a,GAIA16b,lindegren16})\footnote{
http://sci.esa.int/gaia/}.
Among our 340 program stars, 159 have parallaxes reported by both 
surveys, 66 have only Gaia parallaxes, 84 have only Hipparcos parallaxes, 
and just 31 (9\%) have neither.

In Figure~\ref{paracomp} we compare the Gaia and Hipparcos parallaxes for
our RHB stars.
For visual clarity we have chosen plotting axis limits that exclude 
HIP~113747, for which the Hipparcos parallax is negative. 
We also have not extended
the axis limits to include HIP~19302 and HIP~62325, 
which have much larger parallaxes than the rest of the sample 
(Table~\ref{tab-motions}).
The Gaia and Hipparcos parallax values are in excellent agreement for both
of these 
stars: for HIP~19302 ${\pi_{\rm Gaia}}$~ = 22.27~$\pm$~0.48~mas, 
${\pi_{\rm Hip}}$~ = 22.68~$\pm$~0.34~mas, and for HIP~62325 
${\pi_{\rm Gaia}}$~ = 22.41~$\pm$~0.59~mas, 
${\pi_{\rm Hip}}$~ = 22.15~$\pm$~0.35~mas.
In order to avoid clutter in Figure~\ref{paracomp}
we have chosen to display only representative parallax uncertainties. 
We have shown the mean Gaia and Hipparcos parallax uncertainties
for our RHB stars for Hipparcos parallax bins centered at 1, 3, 5, 7, 
and 9~mas, with bin widths of 2~mas.
The uncertainties for Hipparcos parallaxes increase substantially as their 
values decrease; the lower limit for reliable parallaxes in that survey is 
being reached.
For Gaia the opposite trend is seen: the larger parallaxes are more uncertain
than the smaller ones.
This is simply an apparent magnitude effect, as those RHB's with larger
Gaia parallaxes are approaching the brightness limit of the Gaia first 
data release.

Inspection of Figure~\ref{paracomp} shows that for large parallaxes the Gaia 
values agree with the Hipparcos ones, but the Gaia parallaxes are 
systematically higher at the smallest Hipparcos parallaxes.
Arbitrarily dividing the sample into larger and smaller parallax parts, 
for ${\pi_{\rm Hip}}$~$\geq$ 3~mas we have 
$\langle\Delta \pi\rangle$ = $\langle \pi_{\rm Gaia}-\pi_{\rm Hip}\rangle$ = 
0.08~$\pm$~0.09~mas ($\sigma$~=~0.80~mas, 84 stars).
For stars with ${\pi_{\rm Hip}}$~$<$~3~mas, neglecting those with negative Hipparcos
values,  $\langle\Delta \pi\rangle$ = 
1.02~$\pm$~0.12~mas ($\sigma$~=~1.03~mas, 72 stars). There are only eight
stars with ${\pi_{\rm Hip}}$~$<$ 3~mas that do not have Gaia parallaxes.
Moreover, the mean Gaia parallax uncertainties are less than the Hipparcos 
ones at all parallax values. A similar but more comprehensive
discussion has been made by \cite{lindegren16} for a much
larger sample (~$\simeq$~86 000 stars). They adopted a much more rigorous approach and 
compared Hipparcos and Gaia (Gaia DR1) parallaxes for stars in common. Their effort yielded an 
actual parallax uncertainty estimate of 1.218 mas. Given the limits of sample size and 
spectral/evolutionary type of our sample, we regard that uncertainty results as 
roughly comparable, but the \citeauthor{lindegren16} value is probably more representative 
of the true sigma.
Therefore for subsequent computations we will adopt Gaia values when
available, only using the Hipparcos values when necessary.

Calculations of absolute magnitudes were straightforward for targets with 
measured parallaxes: $M_V$ = $-$5log($1/\pi$) + 5 + $V$ $-$ $A_V$,
($\pi$ in arcsec).
About 42\% of our stars lie at distances within 200~pc and nearly 47\% 
are within 500~pc.
Most of these stars should suffer little ISM extinction.
We searched the literature for specific reddening or extinction estimates, 
and in Table~\ref{tab-basic} we quote the literature $A_V$ and $E(B-V)$ values 
when available, or ones computed from one of these quantities using the
standard relation $E(B-V)$~=~$A_V$/3.1.

The target RHB list was composed several years before the first Gaia data
release, and it included RHB candidates that had only rough absolute magnitude 
estimates. Now aided by Gaia parallaxes we have been able to produce a reliable 
CMD of the observed RHB sample.
In Figure~\ref{colormag} we correlate the $B-V$ colors, corrected for 
reddening when available, with derived values of $M_V$.
This figure recreates Figure~1 of \cite{kaempf05}, zoomed in to 
better show the RHB domain in CMD.
Our target stars and those of the (mostly low metallicity) RHB sample of 
\cite{behr03b} are displayed along with stars of the Hipparcos
catalog \citep{vanleeuwen07}.
The polygon delineated with solid lines shows the whole RHB domain defined
by \citeauthor{kaempf05}.  
The dashed vertical line splits the RHB region into ``blue'' and ``red''
parts, with the blue side covering the more easily identified RHB members.
Such stars are substantially bluer than Pop~I solar metallicity RC stars,
occupying the He-burning RHB domain clearly seen in mildly metal-poor 
globular clusters such as NGC~104 (47~Tuc, \citealt{hesser87}), NGC~6366, 
NGC~6624, NGC~6637, and NGC~6838 \citep{rosenberg00a,rosenberg00b}.
The red part of the \citeauthor{kaempf05} RHB domain is bounded on the
cool side by the solid line (or more conservatively the dashed line) that
runs parallel to the RGB branch.
They argue, based on kinematic information, that the great majority of stars
in the area between the slanting dashed and solid lines
should also be RHB stars, not RC stars.
Adopting their RHB definition in $B-V$, $M_V$ space, we find that a large
majority of our targets ought to be RHB stars.

Some of the points that are not in the RHB box in Figure~\ref{colormag} 
appear to be normal MS, SGB, RGB stars, mostly resulting from stellar 
distances that have large uncertainties.
The target RHB stars without parallax information (thus not plotted in
Figure~\ref{colormag}) are on average much fainter than those with 
parallaxes:  they have $\langle V\rangle$~$\simeq$~12.0 while the entire
RHB target sample has $\langle V\rangle$~$\simeq$~8.5 .
If these stars are RHBs ($M_V$~$\sim$~0) and if they suffer no extinction, 
then they should lie at $\langle d\rangle$~$\simeq$~2,500~pc.
But that distance implies substantial reddening if the targets are in
the Galactic disk.
Therefore RHB assignment for such stars will depend entirely on the
spectroscopic model atmosphere parameter derivation to be described
in \S\ref{modelcomp}.

\section{SPACE VELOCITIES AND GALACTIC ORBITAL ELEMENTS}\label{motions}

To calculate the space velocities we used proper motions 
from the Gaia first data release \citep{lindegren16} when available, else from the 
Hipparcos revised catalog \citep{vanleeuwen07}.
With these data and distances taken from the same studies (\S\ref{parmag}), 
and our $RV$ measurements (\S\ref{obsred}), we computed $U$, $V$, and $W$ 
space velocities as described in ASF12.
The basic equations for these velocity components and their uncertainties
were taken from \cite{johnson87}.
These values are with respect to the Sun's motion, and so we shifted them
to local standard of rest (LSR) velocities $U_{\rm{LSR}}$, $V_{\rm{LSR}}$, and $W_{\rm{LSR}}$ 
by adopting the solar peculiar motion with respect to the LSR recommended
by \cite{dehnen98}: $(U, V, W)_\odot$~= ($+$10.00, $+$5.25, $+$7.17)~\kmsec.

In Figure~\ref{toomre2} we summarize these velocities in a so-called Toomre 
diagram \citep{sandage87}, which correlates the Galactic rotational velocity $V$ with the
quadratic sum of the disk radial velocity with respect to the Galactic center
$U$ and the motion out of the plane with respect to the Galactic pole
$W$.  
Dotted lines in the figure denote constant total space velocity 
$V_{\rm{tot}}$~=~$(U_{\rm{LSR}}^2 + V_{\rm{LSR}}^2 + W_{\rm{LSR}}^2)^{1/2}$. 
The solid black line denotes $V_{\rm{tot}}$~=~180~\kmsec.
This is approximately the Galactic velocity dividing line between the disk 
and halo stars (\eg, \citealt{nissen04b}).

Our 309-star sample (the ones with the parallax information) has only 10 with $V_{\rm{tot}}$~$>$~180~\kmsec.
These stars are named in Figure~\ref{toomre2}, along with their [Fe/H]
metallicities derived in this study (\S\ref{modelcomp}).
All but one of these high-velocity stars have [Fe/H]~$<$~$-$0.6, and most
have [Fe/H]~$\lesssim$~$-$1. 
Two more low metallicity stars with $V_{\rm{tot}}$~$<$~180~\kmsec are also 
labeled in the figure.  
These stars, HIP~79518 and HIP~52817 have Galactic disk velocities with 
exceptionally low metallicities. 
They are of secondary interest to our study, 
and will not be featured in the remainder of this paper.

For the rest (vast majority) of our sample, the Toomre diagram with velocity 
limits that encompasses disk members only is given in Figure~\ref{toomre1}.
In ASF12, we assigned stars as ``transition objects''
that had total velocities 50~$<$~$V_{\rm{tot}}$~$<$~70 \kmsec. 
With the advantage of a larger sample in this study, 
we examined our stars in chemo-kinematic plane.
This revealed that about 86\% of our stars in the 
50~$<$~$V_{\rm{tot}}$~$<$~70 \kmsec\ velocity domain have
[$\alpha$/Fe]~$\lesssim$~$+$0.13 and [Fe/H]~$\gtrsim$~$-$0.5 
(see \S\ref{ratios}). 
More than 90\% of the stars with $V_{\rm{tot}}$~$<$~50 \kmsec\ also fall 
into this [Fe/H] and [$\alpha$/Fe] range. 
Therefore, we assign all stars with $V_{\rm{tot}}$~$<$~70 \kmsec\ to the 
kinematic thin disk, absorbing the former transition region in this category.
This choice is in accord with earlier studies, \eg, \cite{loebman11}, 
\cite{nissen09}, and \cite{venn04}.
The stars in the velocity range 70~$<$~$V_{\rm{tot}}$~$<$~180 \kmsec\ are 
thereby designated as thick disk members.
Figure~\ref{toomre1} shows that our sample populates nearly
the whole disk velocity domain, with good representation out to 
$V_{\rm{tot}}$~$\sim$~150~\kmsec.  
There are 215 kinematically-defined thin disk stars, 
80 thick disk stars, 10 halo stars, and 35 objects with no kinematic 
information.\footnote{
The following stars were not included in the 
kinematic calculations:
TYC 23-155-1, because it has a small parallax with an error close to its 
parallax value; BD+54 2710, because it has a parallax error bigger than 
its parallax value; and GSC 01227-00508 and J02221759+5710055, which
have no proper motion information.}

Orbital elements were calculated using a 3-component, 
analytic potential of the Milky Way as in \cite{dinescu99}. 
Uncertainties in these elements were determined via Monte Carlo tests in 
which the initial conditions were varied based on the uncertainties in the 
measured parallaxes, proper motions and radial velocities. 
Upper and lower values of these orbital elements are representative of 
1-sigma errors (\eg, using the 68\% interval centered on the median; see 
\cite{dinescu16}. 
We will discuss the relationships between kinematics, chemical compositions, 
and orbital elements: $z$-component of the angular momentum (L$_{\rm{z}}$); 
total orbital energy (E$_{\rm{tot}}$); eccentricity ($e$); Galactocentric distance (R$_{\rm{m}}$), 
in \S\ref{orbitchem}.

\section{MODEL ATMOSPHERIC PARAMETERS}\label{modelcomp}

\subsection{Initial Estimates}

To derive model atmospheric parameters we used a semi-automated iterative 
software code that was developed by \cite{hollek11}, \cite{roederer14},
and refined for our work by \cite{bocek15,bocek16}.
This approach begins with opening estimates for \teff\ from available 
broad-band photometry and from temperature-sensitive ratios of line depths
in our spectra.  
Initial ``physical'' \logg\ values are calculated from the mean of these 
temperatures and absolute magnitudes either derived from parallax measurements 
or estimated roughly from temperatures and assumptions that the stars 
are evolving along the red giant branch of typical field stars.
Beginning model metallicity values were defined as solar, [M/H]~=~0,
and beginning microturbulent velocities were set as \vmicro~=~1.0~\kmsec\ 
for stars with \teff~$>$~5000~K (probably main sequence 
or subgiant), 1.5~\kmsec\ for stars with 5500~K~$\geq$~\teff~$\geq$~5000~K 
(probably RHB), and 2.0~\kmsec for stars with \teff~$<$~5000~K (RHB or RGB).

Input color temperatures $T_{B-V}$ and $T_{V-K}$ were computed from
the $B$, $V$, $K$, and $E(B-V)$ values of Table~\ref{tab-basic}, assuming
$E(V-K)$~=~2.7$E(B-V)$.
However, we found literature reddening estimates for only 98 (28\%) of 
our 340 program stars.  
Additionally, while 77 of these reddening values are adopted from
\cite{bailerjones11}, the remaining 21 values come from various sources
listed in Table~\ref{tab-basic}.
This constitutes a very heterogeneous set of reddening values, and they 
should be viewed with caution.
In particular, we note that the reddening values for five program stars 
appear to be unreasonably large for their apparent magnitudes:
HIP~34959 ($E(B-V$)~=~0.28, $V$~=~7.87), HIP~38852 (0.36, 8.75), 
HIP~48979 (0.15, 9.64), HIP~52817 (0.25, 8.69), and HIP54858 (0.29, 7.92).
In these cases we neglected reddening corrections when computing input
color temperatures.  
We used the color-\teff\ relationships of \cite{ramirez05} to compute
\teff$_{,B-V}$ and \teff$_{,V-K}$, which are listed in 
Table~\ref{tab-models}.
For a given star, if $|\teff_{,B-V} - \teff_{,V-K}|$~$<$ 200~K we used 
their mean as the color temperature \teff$_{,color}$.
If this difference was beyond 200~K we adopted 
\teff$_{,color}$~=~\teff$_{,V-K}$, as it is less sensitive to 
metallicity than is \teff$_{,B-V}$.

As described in \cite{bocek15}, we used the central depth ratios of up to
14 temperature-sensitive line pairs in the 6200~\AA\ spectral range
to provide \teff$_{,LDR}$ estimates of our RHB candidates.  
The LDR method was developed initially by \cite{gray91}, and expanded by
several authors, most importantly for our work by \cite{biazzo07a,biazzo07b}.
The attraction of the LDR method is that it is insensitive to interstellar 
reddening, and it is a direct spectrum measurement that is not dependent on
a model atmospheric analysis.
The \teff$_{,LDR}$ values from our spectrum depth measurements and 
the \cite{biazzo07a} formulae are listed in Table~\ref{tab-models}.

For the final input \teff$_{\star}$ we again invoked an
empirical rule that if $|\teff_{LDR} - \teff_{color}|$~$<$ 200~K we
used the mean value of these two estimates, but if the difference was
greater than 200~K we used \teff$_{,LDR}$ only because of its freedom from
interstellar reddening uncertainties.

Physical gravities can be calculated in a straightforward manner from:
\begin{eqnarray}                                           
\logg_{phy} = 0.4~(M_{\rm V\star} + BC - M_{\rm Bol\sun}) + \logg_{\sun}  \nonumber
+4~{\rm log} (\frac{\teff_{\star}}{\teff_{\sun}})+ {\rm log} (\frac{{\it m}_{\star}}{{\it m}_{\sun}}).
\end{eqnarray}
where M$_{bol} = 4.75$, log$g$ = 4.44, and \teff\ = 5777 K. 
The bolometric corrections (BC) were estimated using the 
numerical relations provided by \cite{flower96}.

The stellar quantities needed for this equation were not so easy to specify.
First, we lacked information on the masses of our stars, so we arbitrarily
assumed $m_{\star}$~=~1.0~$m_\odot$.
Second, our analyses were completed prior to the first Gaia data release,
so $M_V$ values were available only for about 70\% of our program stars.
For the rest, we assumed initial \logg\ values of 3.7 for \teff~$>$~5500~K
(likely subgiants), 2.3 for \teff~$<$~5000~K (likely red giants), and
3.0 for 5000~K~$<$~\teff~$<$~5700~K (candidate RHB stars).
Finally, for \teff\ we used the mean of the input color and LDR temperatures.

\subsection{Final Parameters}

The initial parameters, and those determined iteratively as 
described below, were used to create interpolated model atmospheres
from the ATLAS grid \citep{kurucz11}\footnote{
http://kurucz.harvard.edu/grids.html}
using software developed by Andy McWilliam and Inese Ivans.

Our semi-automated code uses $EW$'s of \species{Fe}{i}, \species{Fe}{ii},
\species{Ti}{i}, and \species{Ti}{ii} transitions to derive individual
line abundances. The goal
of the model iterations by the code was to determine the model
\teff, logg, \vmicro, and [M/H] that:
\textit{(1)} eliminates or minimizes slopes in the abundance log~$\epsilon$ 
versus excitation energy $\chi$ and versus reduced equivalent width $RW$ 
($\equiv$~log($EW/\lambda$)); 
\textit{(2)} forces agreement between derived log~$\epsilon$ values of the 
two species of Fe and Ti; and 
\textit{(3)} yields derived metallicity [M/H] in agreement
with the assumed model metallicity.
For each iteration on model parameters the abundance trends can be inspected 
in text and graphical form after code completion,
so that a user can judge the quality of the final 
chosen model.
Our use of this code followed those of \cite{bocek15,bocek16} in all aspects.
In particular, in deriving \logg\ from the Fe and Ti ionization balances 
we assigned more weight to Fe (65\%) than to Ti (35\%).
We have also investigated the effects of independent 
Fe and Ti ionization balances by varying the weight given to the Ti ionization balance 
0 and 100 in the \logg\ determinations. For this experiment, we selected
10 RHB candidates that are the good representatives of our entire sample: 
HIP 476, HIP 4197, HIP 8423, HIP 9557, HIP 10872, HIP 23949, HIP 74058, HIP 78990, 
HIP 94598 and HIP 106775. The derived \logg\ values changed by only 0.2 dex on
average, which is well within the uncertainties derived in \S\ref{prevstud}.  

The atomic line lists adopted in this work to compute the model 
atmospheric parameters and species relative abundances are those of
\cite{bocek15,bocek16}; we briefly summarize them here.
The spectral range was about 5200$-$7100~\AA\ to avoid telluric line 
contamination at redder wavelengths and severe line crowding at bluer wavelengths.
We formed lists of relatively unblended lines by using several solar 
atlases (\citealt{moore66}, \citealt{delbouille73}, \citealt{kurucz84}, 
\citealt{wallace11}), the Arcturus spectral atlas \citep{hinkle05}, and the 
interactive database of high-resolution standard star 
spectra SpectroWeb\footnote{
http://alobel.freeshell.org/spectrowebl.html}
\citep{lobel08,lobel11}. 
The model atmospheric parameters of our program RHBs were determined using a
maximum of 70 \species{Fe}{i}, 12 \species{Fe}{ii}, 11 \species{Ti}{i} 
and 6 \species{Ti}{ii} lines; the numbers varied from star to star. 
We used $EW$ limits for both neutral and ionized Fe and Ti lines 
defined by \citeauthor{bocek15}, discarding very weak and very strong lines 
to retain only those in the range
$RW$~$\simeq$ $-$5.8 to $-$4.6 
(equivalent to $EW$~$\simeq$ 10~m\AA\ to 150~m\AA\ at 6500~\AA).
We did not apply these limitations to the other species since they have 
fewer transitions. 
The transition probabilities were taken from single source laboratory-based 
homogeneous studies as much as possible. 
In particular, we used the University of Wisconsin atomic physics group
lab data for \species{Ti}{i} \citep{lawler13}, \species{Ti}{ii} \citep{wood13}, 
and \species{Ni}{i} \citep{wood14b}. 
Unfortunately the fundamental metallicity-defining element Fe has no single
comprehensive lab study for either its neutral or ionized species.
Therefore we adopted several sources that together should yield reliable
species abundances.  
For \species{Fe}{i}, those are \cite{obrian91} (lines with excitation energies
$\chi$~$<$~2.1~eV) and \cite{denhartog14}, \cite{ruffoni14} (lines with
$\chi$~$\geq$~2.1~eV).
For \species{Fe}{ii} (lacking any recent detailed lab work), we used 
$gf$-values from the NIST Atomic Spectra Database\footnote{
https://www.nist.gov/pml/atomic-spectra-database}.
For $gf$ values of remaining species \species{Si}{i}, \species{Ca}{i},  
\species{Cr}{i} and \species{Cr}{ii} we have used several sources, which are
given in Table~4 of both \cite{bocek15} and \cite{bocek16}.

In Figure~\ref{teff3guess} we compare the final spectroscopic temperatures
\teff$_{,sp}$ to the initial LDR and photometric estimates.
In panel (a) the correlation with \teff$_{,LDR}$ is linear with small scatter.
For the 270 stars with \teff$_{,sp}$~$<$~5200~K, a simple mean offset is
$\langle \teff_{,LDR} - \teff_{,sp}\rangle$ = 108~$\pm$~7 ($\sigma$~=~124).
Above this temperature the \teff$_{,LDR}$ values asymptotically approach 
5300~K while the \teff$_{,sp}$ ones continue to rise.
The \cite{gray91} and \cite{biazzo07a} LDR formulae predominantly use the 
depths of low excitation \species{V}{i} lines as the numerators of their 
ratios. 
At the $R$ and $S/N$ of our spectra, these \species{V}{i} lines become very 
weak at warmer \teff\ values and the \teff\ formulae tend toward single values.
ASF12 included the LDR ratios of \cite{strassmeier00}, which used species
other than \species{V}{i}, for stars with \teff$_{,sp}$~$\geq$~5500~K
(see their Figure~3).
However, the comparison between spectroscopic and LDR temperatures for
the \citeauthor{biazzo07a} and \citeauthor{strassmeier00} relations were
offset by roughly 300~K.
Since stars with \teff~$\gtrsim$~5500~K were not of prime interest to our
RHB work, we have not pursued this issue further.

At the lower temperature and higher metallicity end of the
sample, our chosen Fe and Ti lines have strengths large enough that all 
lie on the flat or damping parts of the curve-of-growth.
For these stars, it is very difficult to determine independent values of 
\teff\ and \vmicro. 
Therefore, for a few of these stars we adopted a fixed 
value of \vmicro~between 1.2 and 1.6~\kmsec, and varied only \teff, \logg, 
and [Fe/H] during the model iterations. 
We also adopted fixed \vmicro\ values between 0.5 and 1.2~\kmsec\ for about 
30 stars with \logg~$>$~3.5.

In panels (b) and (c) we compare spectroscopic temperatures
with $(V-K)$ and $(B-V)$ photometric temperatures.
Qualitatively the correlations show good agreement with reasonably
small scatter for stars with \teff$_{,sp}$~$\lesssim$~5200~K,
but significant scatter appears for warmer stars.
The following cautions should be kept in mind here.
The photometry is heterogeneous, taken from SIMBAD and ultimately
based on a variety of original sources.
The reddening estimates exist only for some of our 
RHB stars, and those reddening values have various uncertainties.
Finally, as discussed in \cite{ramirez05} $(B-V)$ photometric temperatures 
are more dependent on \logg\ and metallicity than are $(V-K)$ temperatures. 
This creates more scatter among the $(B-V)$ temperatures, especially at  
temperatures $\gtrsim$~5200~K. 

Thus it is not surprising that there are relatively weak correlations between
spectroscopic and photometric temperatures.

In Figure~\ref{gphysgspec} we compare the input gravities \logg$_{phys}$ 
computed with the above formula, and those derived from our spectroscopic
analysis.
On average the correlation is good, albeit with large star-to-star
scatter:  $\langle \logg_{phys} - \logg_{spec}\rangle$ = $+$0.05~$\pm$~0.03 
($\sigma$~=~0.51, 318 stars).  
Our physical gravity estimates depend linearly on assumed
stellar masses, and in the figure we indicate how these gravities  would change
if the masses were doubled.

The complete set of derived model atmospheric parameters
for our program stars are listed in Table~\ref{tab-models}.
Using the final \teff\ and \logg\ values, we have constructed an HR diagram 
in spectroscopic units in Figure~\ref{tefflogg}.
The gray-shaded regions of this plot, \teff~$>$~5500~K, \teff~$<$~4500~K, 
and/or \logg~$>$ 3.5, represent the parameter space of program stars that 
are unlikely to contain true RHB members.
The color range of our program stars inside the 
RHB domain defined by \cite{kaempf05} is $(B-V)$~$\simeq$~0.6$-$1.0
(Figure~\ref{colormag}).
This color interval corresponds to
 a temperature range 
\teff~$\sim$~4800$-$5800~K, using the color-temperature calibrations of 
\cite{ramirez05}; see their Figure~3.  
Our sample has only seven stars with 5500~K~$<$~\teff~$<$~5800~K and 
\logg~$<$~3.5; just one of these has \logg~$<$~3.0.
Proof of chemically-evolved status \citep{afsar18}
for these stars would be difficult because they will have generally very
weak/absent molecular bands and [\species{O}{i}] lines.
Therefore they are not included in the RHB domain as defined here.
We do retain stars in the 
4500~K~$<$~\teff~$<$~4800~K range for now, and will
consider their RHB/RGB status in the next paper.

ASF12 defined RHB stars through a combination of position in the \teff-\logg\
plane and proof of internal CN-cycle H-fusion and mixing, as
evidenced in low surface carbon isotopic ratios, \carbiso~$<$~30.
They did so because stars with masses $m$~$\sim$~2.5$-$4.5~$m_\odot$ pass
through the RHB domain during their SGB evolution, prior to their completion
of the 1$^{st}$ dredge-up phase.
A large majority of the ASF12 RHB candidates with temperatures between 
4800~K and 5500~K proved to be mixed stars.
Of the 45 stars in this \teff\ range with \logg~$<$~3.5, nine had 
\carbiso~$\gtrsim$~30 or undetermined (20\%); these stars had essentially 
undetectable $^{13}$CN features in their spectra.
The isotopic ratios derived for the remaining 36 stars (80\%) showed clear 
evidence for internal evolution: 15 of them had \carbiso~=~20-29, 14 had
\carbiso~=~10-19, and 7 had \carbiso~$<$~10.
Therefore we expect our program stars with in this (\teff, \logg) domain of
Figure~\ref{tefflogg} to be dominated by evolved stars.
We adopt the unshaded region of Figure~\ref{tefflogg}, 
4500~K~$\leq$~\teff~$\leq$~5500~K and \logg~$\leq$~3.5, as the ``RHB domain''.
This is a generous definition, and includes stars with \logg\ values between
3.0 and 3.5 that may well turn out to be subgiants, and stars with 
\teff~$\lesssim$~4800~K that may be an admixture of RC and RGB stars.
More definitive RHB membership statements about individual stars will
require the CNO analysis in \cite{afsar18}.

\subsection{Parameter Uncertainties and Comparison with Previous 
            Studies}\label{prevstud}

To investigate the internal uncertainties in atmospheric parameters,
we selected two stars that are good representatives of our entire sample; 
HIP 4197 and HIP 10872. 
We ran a series of analyses by varying the stellar 
atmospheric parameters in steps of 50 K for \teff, 0.05 dex for \logg\ and 
0.05 \kmsec\ for \vmicro\ on the spectral data of both stars. 
Individual uncertainties were estimated by varying one parameter at a time 
while others were fixed during the each analysis.
To determine the uncertainty in \teff\ we searched for the temperature 
excursion that creates a $\pm$1$\sigma$ scatter in the 
abundance difference between low- and high-excitation \species{Fe}{i} lines. 
Applying this method resulted in estimated \teff\ 
uncertainties of $\sim$150~K for both stars.
The same method was applied to determine the uncertainties in \logg\ 
and \vmicro.
The typical average uncertainties in these parameters were found to be 
$\sim$0.25 dex and $\sim$0.25 \kmsec, respectively. 
We have also investigated the effects of atmospheric 
parameter uncertainties on the elemental abundances of both HIP 4197 and 
HIP 10872 and listed the mean sensitivities in Table~\ref{tab-uncert}.
The results indicate that effective temperature uncertainties contribute 
most to the abundance uncertainties of neutral species, while surface gravity 
uncertainties mainly affect those of the ionized species. 
However, this means that the sensitivities of abundance ratios [X/Fe] are not 
large (see \S\ref{ratios}).

External (scale) uncertainties can be estimated 
by comparing our results with the previous studies that have stars 
in common with our sample.
First we comment on the parameters newly determined 
in this study for the stars previously published in ASF12.
There are 45 such stars, counting the few with double analyses here as separate
objects.
Forming mean differences for an atmsopheric quantity $X$ in the sense 
$\langle\Delta X\rangle$~=~$\langle X_{ASF12} - X_{new}\rangle$, we find
$\langle\Delta \teff\rangle$~= $+$23~$\pm$~12~K ($\sigma$~=~87~K),
$\langle\Delta \logg\rangle$~= $+$0.17~$\pm$~0.04 ($\sigma$~=~0.30),
$\langle\Delta \vmicro\rangle$~= $-$0.07~$\pm$~0.02 ($\sigma$~=~0.16), and
$\langle\Delta {\rm [Fe/H]}\rangle$~= $+$0.03~$\pm$~0.02 ($\sigma$~=~0.11).
In ASF12 we made use of only neutral and ionized 
species of Fe for model atmosphere analysis, while the present
work takes two species of Fe and Ti into account for the analysis.
The effect of updated $gf$-values on model atmosphere calculations should 
also be considered when assessing the difference between the old and 
newly determined parameters.
Our new analyses are in good accord with those of ASF12.

The largest compilation of stellar parameters relevant to our work
is the PASTEL catalog \citep{soubiran16}.
There are many more entries in this catalog for MS, SGB, and RGB stars
than for the comparatively rare RHB stars.
A non-exhaustive search through this database found 225 entries for
88 of our program stars.  
Here we opted to treat multiple PASTEL
entries from different literature sources for individual stars as 
separate entities.
In Figure~\ref{pastelcomp} we correlate 
that catalog's \teff, \logg, and [Fe/H] values with 
those determined in this work.
Taking differences in the sense PASTEL\ entry $minus$ this\ study,
we find good agreement for effective temperature and metallicity: 
$\langle\Delta\teff\rangle$~= $+$45~$\pm$~7~K ($\sigma$~=~102~K), and
$\langle\Delta {\rm [Fe/H]}\rangle$~= $+$0.03~$\pm$~0.01 ($\sigma$~=~0.12).
Our gravities are typically smaller than those from the literature:
$\langle\Delta\logg\rangle$~= $+$0.29~$\pm$~0.02 ($\sigma$~=~0.34),
but the star-to-star scatter clearly is large.
The uncertainty in gravity does not appear to affect the metallicity
comparison, and our \logg\ values are on an internally-consistent system,
whereas the PASTEL entires reflect the results of scattered literature
sources.

\cite{luck15} (and references therein)
reported model atmosphere parameters and abundances for 1133 F, G, and K 
(3800~K~$\gtrsim$~\teff~$\gtrsim$~7800~K) luminosity class III stars.
We have 48 stars in common with their more general evolved star survey.
Figure~\ref{luckcomp} panels (a)-(c) illustrate the excellent agreement 
between our derived values of \teff, \logg, [Fe/H] and theirs.
Defining differences in the sense 
$\Delta$X $\equiv$ X$_{luck15}$~$-$ X$_{this\ study}$, we find:
$\Delta$\teff~= $+$15~$\pm$~13~K ($\sigma$~= 88~K);
$\Delta$\logg~= $+$0.01~$\pm$~0.03 ($\sigma$~= 0.24); and
$\Delta$[Fe/H]~= $+$0.04~$\pm$~0.01 ($\sigma$~= 0.09). 
The mean values and scatters are entirely within the uncertainties in these
quantities in the two studies.
We will comment on the [$\alpha$/Fe] values of Figure~\ref{luckcomp}
panel (d) below.

\section{DERIVED ABUNDANCES}\label{abunds}

\subsection{Metallicities}\label{metal}

Our program stars span the derived metallicity range
$-$2.0~$\lesssim$~[Fe/H]~$\lesssim$~$+$0.4, but only two stars
have [Fe/H]~$<$~$-$1.5:
HIP~38852 ([Fe/H]~=~$-$2.05) and HIP~79518 ($-$1.89).
Although we have a large stellar sample, our target 
selection was accomplished primarily with temperature and luminosity criteria;
no attempt was made to achieve statistical completeness of the disk
metallicity domain.
For program stars with measured parallaxes 
(Table~\ref{tab-motions}), only 12 lie at distances greater than 1000~pc.
Although some of the 31 stars without parallax data may be at large distances,
at most this is about a 10\% effect; our stellar sample is relatively local.
Thus major statements about overall Galactic disk populations are not 
warranted.
With that caution in mind, a few general conclusions can be drawn about 
the metallicity distribution function of our sample.

In Figure~\ref{fehisto} we show metallicity histograms 
of our stars with [Fe/H]~$\geq$~$-$1.5, excluding the two 
low-metallicity stars which are of less interest here.
In panel (a) histograms are displayed for all stars and for just the ones 
in the RHB domain 
as defined in \S\ref{modelcomp} and depicted in Figure~\ref{tefflogg}.
These two histograms are very similar, thus are sampling the same Galactic 
disk metallicity distribution function.
There is a significant tail of stars with [Fe/H]~$\lesssim$~$-$0.5, indicating 
the presence of a substantial thick disk component to our sample.
In this panel we mark the locations of the median and mean values,
[Fe/H]~=~$-$0.20 and $-$0.28, respectively.
These are significantly offset from the solar metallicity.

Our field RHB metallicity distribution appears to be
consistent with those of some other spectroscopic Galactic disk surveys.
Here we consider two examples.
\cite{bensby14} conducted an abundance study of 714 F- and G-type dwarf and
subgiant stars that were kinematically chosen to represent disk and halo
components of the Galaxy.  
Inspection of their Figure~2 suggests that the metallicity distribution 
of their stars in kinematic thin disk, thick disk, and halo bins 
do not resemble that of our program stars.
However, selecting the \citeauthor{bensby14} stars with thick disk to
thin disk kinematic probability ratios $TD/D < 2$ (the stars of panels (c) 
and (d) of their Figure~2) produces the metallicity distribution shown
in panel (b) of Figure~\ref{fehisto}.
This distribution is broadly similar to ours.
A systematic shift in our metallicity scale by $+$0.1~dex would produce
essential agreement between the peaks of two distributions for stars with 
[Fe/H]~$\gtrsim$~$-$0.2~dex, but the metallicity widths are clearly a bit different 
$-$ our distribution in [Fe/H] appears to be narrower than that of  \citeauthor{bensby14}.

The APOGEE near-IR spectroscopic survey \citep{holtzman15,majewski16}
is much more statistically robust, and inspection of Table~2 and Figure~5 
of \cite{hayden15} suggests that our average and peak metallicities
correspond to stars with Galactocentric radius $R$~$\sim$~10$-$11~kpc
for heights above the Galactic plane inside $|z|$~$<$~0.5~kpc, and
$R$~$\sim$~6$-$10~kpc for 0.5~$<$~$|z|$~$<$~1.0~kpc.  
These distance parameters cover a Galactic
volume not very much different than that containing our program stars.
Given the lack of cross-calibration between our metallicity scale
and that of APOGEE, the concordance is reasonable.

\subsection{Abundance Ratios}\label{ratios}

In this paper we report abundances for the $\alpha$ elements Si and Ca,
the $\alpha$-like element Ti\footnote{
True $\alpha$ elements are those whose major 
naturally-occurring isotopes are composed of 
multiples of \iso{4}{He} nuclei; these include
the observable elements Mg, Si, S, and Ca. 
The Ti abundance is dominated by \iso{48}{Ti} 
(73\% in the solar system),
and the $\alpha$ from \iso{44}{Ti} is unstable with a short half-life.
Since the [Ti/Fe] ratio exhibits the same general trend with metallicity
as does the real $\alpha$ elements, it often is given the same element
group name.}
and Fe-group elements Cr and Ni.
Both neutral and ionized species have been 
investigated for Ti and Cr.

Abundance ratios [X/Fe] for all program stars are listed 
in Table~\ref{tab-abunds} and plotted as functions of [Fe/H] metallicities in 
Figure~\ref{allelements}. These ratios have been formed from same-species 
abundances, that is [\species{X}{i}/\species{Fe}{i}] and [\species{X}{ii}/\species{Fe}{ii}]. 
Therefore, as mentioned in \S\ref{prevstud}, the sensitivities of these ratios to 
to the uncertainties in atmospheric parameters 
are much smaller than the absolute uncertainties in the individual species abundances.

For comparison with our results, we also have plotted the metallicities and
abundance ratios of three large-sample surveys of Galactic disk main sequence
and early subgiant field stars published by \cite{reddy03,reddy06} and 
\cite{bensby14}.

In general the abundance ratios of our program stars follow the trends
with metallicity established by these higher-gravity field stars.
From Figure~\ref{allelements} we note that:
\begin{enumerate}
\item $\alpha$ elements of left-hand panels (a)$-$(d) show the familiar
increase in relative abundance ratios with decreasing [Fe/H] metallicity, 
reaching [X/Fe]~$\simeq$~$+$0.25 with $\pm$0.15 scatter.
\item The two species of Ti are in reasonable accord as expected from 
their roles in assisting with our \logg\ determinations; 
$\langle$\species{Ti}{ii}$-$\species{Ti}{i}$\rangle$~=
$+$0.06 ($\sigma$~=~0.06).
\item Si exhibits much star-to-star scatter (panel a).
We repeated the analysis for a few stars that show considerable scattering 
in Si abundance. 
A common feature in six of these stars with [Fe/H]~$\gtrsim$~0 and 
[Si/Fe]$\sim$0.35 dex is that the lines used for the model atmosphere analysis 
have a very narrow RW range that requires a fixed microturbulent velocity for 
the solution. 
Therefore, these results should be interpreted with caution. 
A further concern is the effective temperature dependence of Si.
As seen in Figure~\ref{Teff_Si}, [Si/Fe] increases as the \teff\ decreases. 
Investigation of the metallicity dependency of this relation showed no obvious 
connection to the metallicity, in other words, whatever the metallicity is, 
the relative Si abundance seems to increase depending on effective 
temperature.\footnote{
A similar effect for very metal-poor stars has been found 
in Si abundances derived from the \species{Si}{i} 3905~\AA\ line
(not employed here); see, \eg, \cite{sneden08b}, and references therein}.
Underlying causes of this behavior need further investigation.
\item The two Cr species have a mean abundance offset
(panels e, f):
$\langle$\species{Cr}{ii}$-$\species{Cr}{i}$\rangle$~=$+$0.05.
However, many \species{Cr}{ii} abundances are anomalously large, leading
to $\sigma$~=~0.31, a scatter that overwhelms the mean offset.
For \species{Cr}{ii} we have only five transitions 
(4588.20, 4592.05, 5237.32, 5305.87 and 5334.86).  
Our adopted transition probabilities from \cite{bocek15} have recently
been revised by \cite{lawler17}.
For our lines the scatter old-new in log~$gf$ values is high but the
mean is essentially the same.
The 4588 and 4592 lines are in a crowded spectral region; blending with
other transition(s) may be of concern.
For most of the RHBs there are only one or two \species{Cr}{ii} lines to be 
trusted for the abundance.  
Moreover, in Table~\ref{tab-abunds} there are RHBs whose \species{Cr}{ii} 
abundances are based on only one line. 
For all of these reason the overall star-to-star scatter in \species{Cr}{ii}
abundances is higher than in \species{Cr}{i}; caution is warranted in
interpreting  panel (f) of Figure~\ref{allelements}.
\end{enumerate}

Further insight is gained by forming mean abundances of the Fe group and
$\alpha$ elements.
In panel (a) of Figure~\ref{means} we show the mean [X/Fe] values of the
three Fe-group species. 
The trend with metallicity is nearly flat, with little star-to-star variation.
Defining [Fe group/Fe]~= $\langle$[\species{Cr}{i}/Fe],[\species{Cr}{ii}/Fe],
[\species{Ni}{i}/Fe]$\rangle$, 
for the whole sample, $\langle$[Fe group/Fe]$\rangle$~=~0.00 with 
$\sigma$~=~0.05.
These values are, as expected, consistent with solar Fe group abundance 
ratios and star-to-star scatter dominated by observational/analytical 
uncertainties.

In panel (b) of Figure~\ref{means} we plot [$\alpha$/Fe] 
(the simple mean of [\species{Si}{i}/Fe], [\species{Ca}{i}/Fe], 
[\species{Ti}{i}/Fe], [\species{Ti}{ii}/Fe]) versus [Fe/H].
A pale red line has been drawn at 
[$\alpha$/Fe]~=~$+$0.13 with a width $\delta$[$\alpha$/Fe]~=~0.04.
This line is a reasonable estimate of the division between
$\alpha$-solar and $\alpha$-rich stars in our sample. 
This separation has been suggested by several authors and has been
studied extensively in the very large APOGEE survey (\eg, \citealt{martig15} 
and references therein).
In our sample the $\alpha$-rich group occurs at lower metallicities, 
[Fe/H]~$\lesssim$~$-$0.3 and has an average [$\alpha$/Fe]$~\simeq$~$+$0.25.
The growth in [$\alpha$/Fe] with decreasing [Fe/H] reaches a plateau at 
[Fe/H]$\simeq$~$-$0.5, in accord with so-called the ``knee" defined in
previous studies (\eg, \citealt{bensby03,reddy06,liu12,bensby17}). 
At lower metallicities $\langle$[$\alpha$/Fe]$\rangle$ remains nearly
constant, and at higher metallicities it declines to about +0.05.  
The average metallicity of the group with solar [$\alpha$/Fe] values occurs 
at about [Fe/H]$~\simeq$~$-$0.1. 
This is all in accord with previous results from low luminosity main sequence 
and subgiant stars (\eg, \citealt{reddy06,anders14,martig15,bensby17}).
Additionally, all but one of our 48 program stars also 
investigated by \cite{luck15} (see \S\ref{prevstud}) fall into the
metallicity range [Fe/H]~$\gtrsim$~$-$0.7 (the lone exception has 
[Fe/H]~=~$-$1.43).
Inspection of panel (d) of Figure~\ref{luckcomp} shows that the agreement
in [$\alpha$/Fe] between the two studies is excellent: 
$\Delta$[$\alpha$/Fe]~= $+$0.01~$\pm$~0.00 ($\sigma$~= 0.03).

\section{CORRELATIONS BETWEEN KINEMATICS AND CHEMICAL COMPOSITIONS}\label{corrkinchem}

Understanding the relation between chemical compositions 
and kinematics of stars is essential to better constrain Galactic 
evolutionary scenarios. 
In Figure~\ref{mean_kin}, we present a version of Figure~\ref{means}(b) with 
the symbols coded in thin disk, thick disk, and halo velocity groups 
defined as in Figures~\ref{toomre2} and \ref{toomre1}.
The lower metallicity $\alpha$-rich group 
([$\alpha$/Fe]~$>$~$+$0.13) is dominated by thick disk stars. 
The very high velocity stars presumed to be members of the Galactic halo 
have [$\alpha$/Fe] ratios generally consistent with those of $\alpha$-rich
thick disk stars. 
Higher metallicity $\alpha$-solar group has nearly all of the thin disk stars, 
but also has a significant amount of thick disk stars with nearly circular 
orbits (see \S\ref{orbitchem}). 
Since the chemical separation seems more discernible than kinematic
separation, here we prefer to use the term 
``chemically thin disk ($\alpha$-solar) / chemically thick disk ($\alpha$-rich)''.

In Figure~\ref{FeH_alpha_kin}, we plot the $UVW$ 
velocity components as functions of [Fe/H] and [$\alpha$/Fe]. 
At a metallicity of [Fe/H]~$\lesssim$~$-$0.3, all velocity components begin
to substantially increase their dispersions, as 
discussed in, \eg, 
\cite{edvardsson93}, \cite{reddy06}, \cite{bensby11} and \cite{anders14}. 
In [$\alpha$/Fe], this velocity-dispersion increase appears at about 
$+$0.13~dex. 
Of the three velocity components, $W_{\rm{LSR}}$ component shows this 
transition most distinctly in both [Fe/H] and [$\alpha$/Fe] (panels (c) 
and (f), respectively).
Specifically, the transition is from a dispersion of about 
{$|W_{\rm{LSR}}|$~$\simeq$~40 \kmsec 
at [$\alpha$/Fe]~$\simeq$~$+$0.13 to 80 \kmsec at 
[$\alpha$/Fe] $>$~$+$0.13. 
This sharp change in the velocity dispersion represents the transition 
between the kinematically cold thin disk to the kinematically hotter 
thick disk, and has been seen before, \eg, in 
\cite{reddy06} and \cite{brook12}.

\section{ORBITAL CHARACTERISTICS AND CHEMICAL COMPOSITIONS}\label{orbitchem}

In Figure~\ref{FeHalpW_ecc} we investigate the 
distribution of metallicities, relative $\alpha$ abundances, and 
$W_{\rm{LSR}}$ velocity components as functions of Galactic orbital 
eccentricities, $e$.
As seen in panel (a), [Fe/H] has a strong trend with eccentricity
kinematically thick disk stars, for those with 
$e$~$>$~0.25, as previously discussed by, \eg, \cite{lee11} and \cite{boeche13}.
The [$\alpha$/Fe] ratios also correlate with eccentricity (panel (b) of the
figure) but there is less distinct grouping $-$ a number of thin disk stars are
$\alpha$-rich and many thick disk stars are $\alpha$-solar 
([$\alpha$/Fe]~$\lesssim$~+0.13, $W_{\rm{LSR}}$~$\lesssim$~40 \kmsec).
Finally, the dispersion in $W_{\rm{LSR}}$ space velocities clearly is larger
for the thick disk and halo stars than those of the thin disk (panel (c)
of the figure).
In short, these three observable aspects of stars in our RHB sample 
all correlate with derived orbital eccentricities, and yield a 
set of stellar population characteristics that is consistent with that
determined from main sequence stars in previous surveys 
(as cited above, particularly in \S\ref{ratios}).

Further insight can be gained from correlations between
eccentricity, [Fe/H] and [$\alpha$/Fe] with Galactocentric distance,
as displayed in Figure~\ref{Rm_e_alp_Fe}. 
We define a mean Galactocentric distance as 
R$_{\rm{m}}$~=~(R$_{\rm{pericenter}}$ + R$_{\rm{apocenter}}$)/2.
The low eccentricity, $\alpha$-solar, high metallicity stars are mostly 
located within the solar radius zone (7 kpc~$<$~R$_{\rm{m}}$~$<$~9 kpc).
However, our RHB stars with thick disk and halo 
eccentricities (panel (a) of the figure) generally 
lie at radii smaller than that of the solar circle}; most have
5 kpc~$<$~R$_{\rm{m}}$~$<$~7 kpc.
The separation in R$_{\rm{m}}$ becomes more distinct if the 
higher-metallicity, lower [$\alpha$/Fe] stars are grouped with the thin 
disk members.
The overwhelming number of thick disk RHB stars with 
[$\alpha$/Fe]~$\gtrsim$~$+$0.2, and [Fe/H]~$\lesssim$~$-$0.3 have relatively
small R$_{\rm{m}}$ values.

Although our sample along R$_{\rm{m}}$ is not complete, 
it still resembles the findings presented by, e.g., \cite{edvardsson93}, 
\cite{anders14} and \cite{hayden15}, where they discuss the 
[$\alpha$/Fe]$-$[Fe/H] relation over a large Galactocentric distance. 
In all cases, two distinct $\alpha$-groups appear with a mean 
[$\alpha$/Fe]~$\simeq$~0.25 dex for the inner-group (R$_{\rm{m}}$~$<$~7 kpc) 
stars. 
The mean $\alpha$ abundance is $\langle$[$\alpha$/Fe]$\rangle$~$\simeq$~0.1 dex 
for the outer-group (R$_{\rm{m}}$~$>$~8.0 kpc) stars.
In our sample, inner-stars are mostly composed of kinematically thick 
disk stars, as discussed above, while the outer-group contains all 
kinematic classes. 
Interestingly, four of the nine stars kinematically classified as halo 
members are located the closest to the Galactic center in our entire sample. 
Their [$\alpha$/Fe] values are all high (around 0.3), while their [Fe/H] 
metallicities vary substantially.

With Figure~\ref{alpharp_Evt_Lz}, we investigate the 
total orbital energy (E$_{\rm{tot}}$) and the orbital angular momentum (L$_{\rm{z}}$), 
i.e., the integrals of motion in an axisymmetric potential. 
In panel (a) of the figure we plot the E$_{\rm{tot}}$
as a function of the L$_{\rm{z}}$. 
Grey triangles represent $\alpha$-rich ([$\alpha$/Fe]~$>$~$+$0.13) stars, 
while red dots represent the $\alpha$-solar stars in our sample. 
$\alpha$-solar stars occupy the domain of rotationally supported orbits, 
predominantly the thin disk.\footnote{
The lone exception is a star that is on a retrograde 
orbit of relatively low energy. 
With an [Fe/H] = $-$0.6, this high-velocity star is likely accreted 
from a satellite similar to the present-day dwarf spheroidal (dSph) galaxies 
orbiting the Milky Way.
This star fits the dSph [$\alpha$/Fe] versus [Fe/H] pattern 
(e.g., \citealt{geis07}), more specifically a Sagittarius dwarf-like system.}
The $\alpha$-rich stars occupy the whole kinematic domain, from the thin/thick 
disk to halo. 
Among these stars, the most energetic ones are on retrograde orbits.

However, most of the $\alpha$-rich stars have lower E$_{\rm{tot}}$ with less 
L$_{\rm{z}}$, indicating that they mostly spend their time in the inner 
regions of the Galaxy and have less disk-like rotation than the 
$\alpha$-solar stars. 
This result is in accord with our conclusions from the velocity analysis. 
We also note a tight group of five stars in this plot. 
They have nearly zero angular momentum and very low orbital energy, 
practically the lowest in our sample; we will further investigate these below. 
In Figure~\ref{alpharp_Evt_Lz} panels (b) and (c), 
we show the orbital eccentricity, $e$, as a function of E$_{\rm{tot}}$ 
and L$_{\rm{z}}$, respectively (Galactic orbital parameters are listed in 
Table~\ref{orbpar}). 
$\alpha$-solar stars have moderate energies with mostly disk-like and 
nearly circular ($e$~$<$~0.3) orbits, though there are a few exceptions. 
The most notable is the star with the highest eccentricity, $e$~$\simeq$~0.8;
this is the same star discussed above, which is on a retrograde orbit 
(see also panels (a) and (c)), and likely is an accreted star. 
$\alpha$-rich stars show all the range of eccentricities. 
Two high-velocity $\alpha$-rich stars that appear on retrograde orbits 
(panel (c)) draw attention with their thick-disk-like eccentricities.
The most energetic stars are also highly eccentric; however, 
the lowest energy ones are also highly eccentric. 
These latter ones are the same group of five stars noted before, with 
nearly zero angular momentum (panels (a) and (c)).
They are practically on radial orbits, and spending their time in the 
inner regions of the Galaxy. 
Their [$\alpha$/Fe] ratios range from $+$0.23 to $+$0.30 with an average 
of $+$0.28 dex ($\sigma$~=~0.03)
while [Fe/H] varies from $-$2.05 to $-$0.49. 
Their Ni and Cr abundances are nearly solar. 
The [Fe/H] and [$\alpha$/Fe] ranges of these halo stars with high-eccentricities and 
low-energies quite similar to the stars presented in, e.g., \citealt{geis07}, 
in which they discuss the Galactic halo formation (see  their Figure 12).  
We will further investigate the profile of [$\alpha$/Fe]~$-$~[Fe/H] relation 
for these stars, together with other element abundances in our future work \cite{afsar18}.

\section{SUMMARY AND DISCUSSION}\label{discuss}

We have conducted a high-resolution spectroscopic survey 
of 340 field red horizontal branch candidates.
In this first paper on survey results we have focused on the chemo-kinematics 
of targets, deriving abundances from only those $\alpha$ and Fe-group element
transitions that can be easily analyzed from EW measurements.            

\begin{figure}
\epsscale{1.00}
\plotone{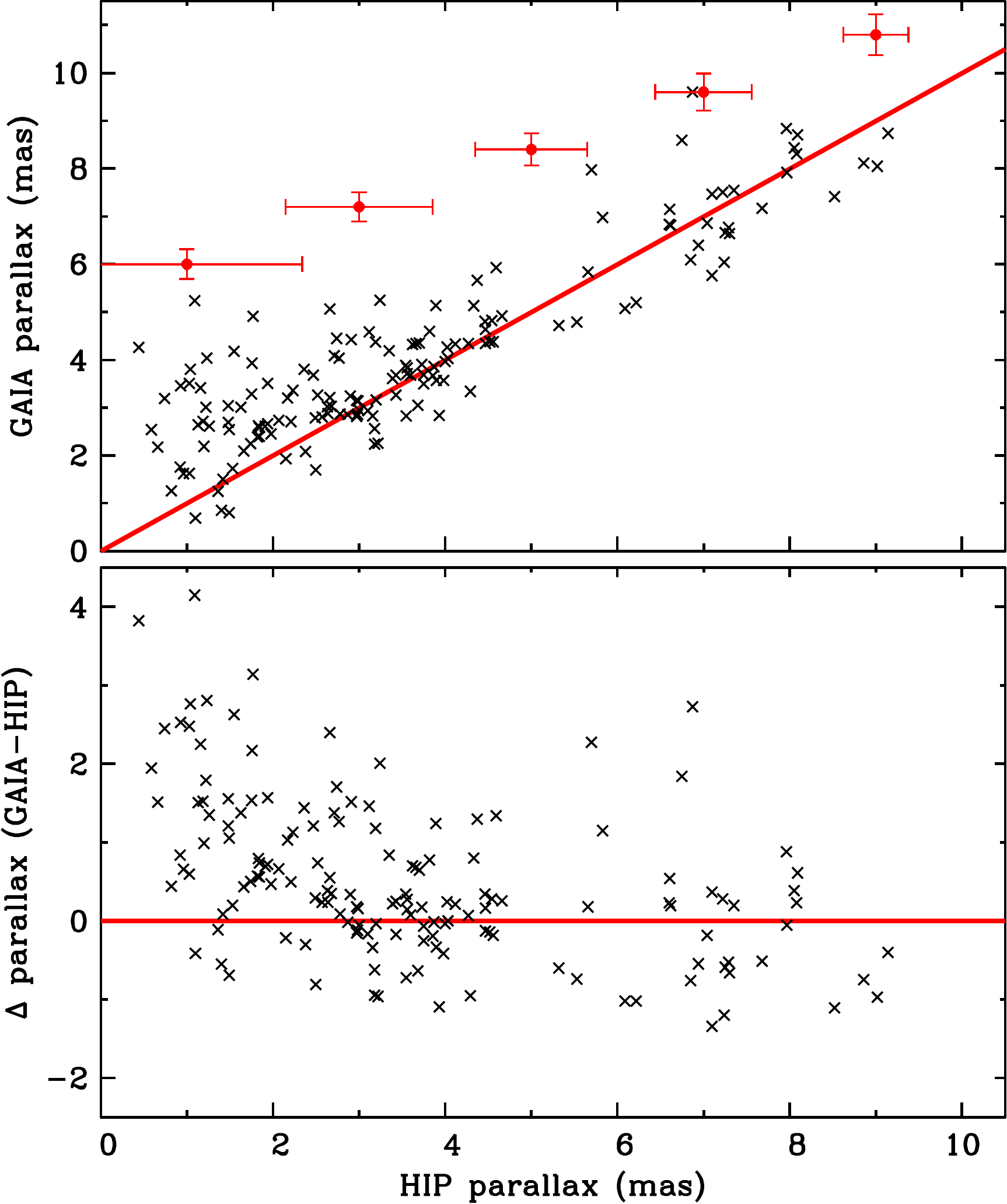}
\begin{center}
\end{center}
\caption{\label{paracomp}
\footnotesize
    Comparison of parallaxes obtained by Hipparcos \citet{vanleeuwen07}
    and Gaia \citet{lindegren16}.
    The red lines denote perfect agreement between the two datasets.
    In the top panel mean uncertainties in the Hipparcos and Gaia parallaxes
    are shown for the Hipparcos values in bins 2~mas wide, centered at
    1, 3, 5, 7, and 9~mas.
}
\end{figure}

\begin{figure}
\epsscale{1.00}                                               
\plotone{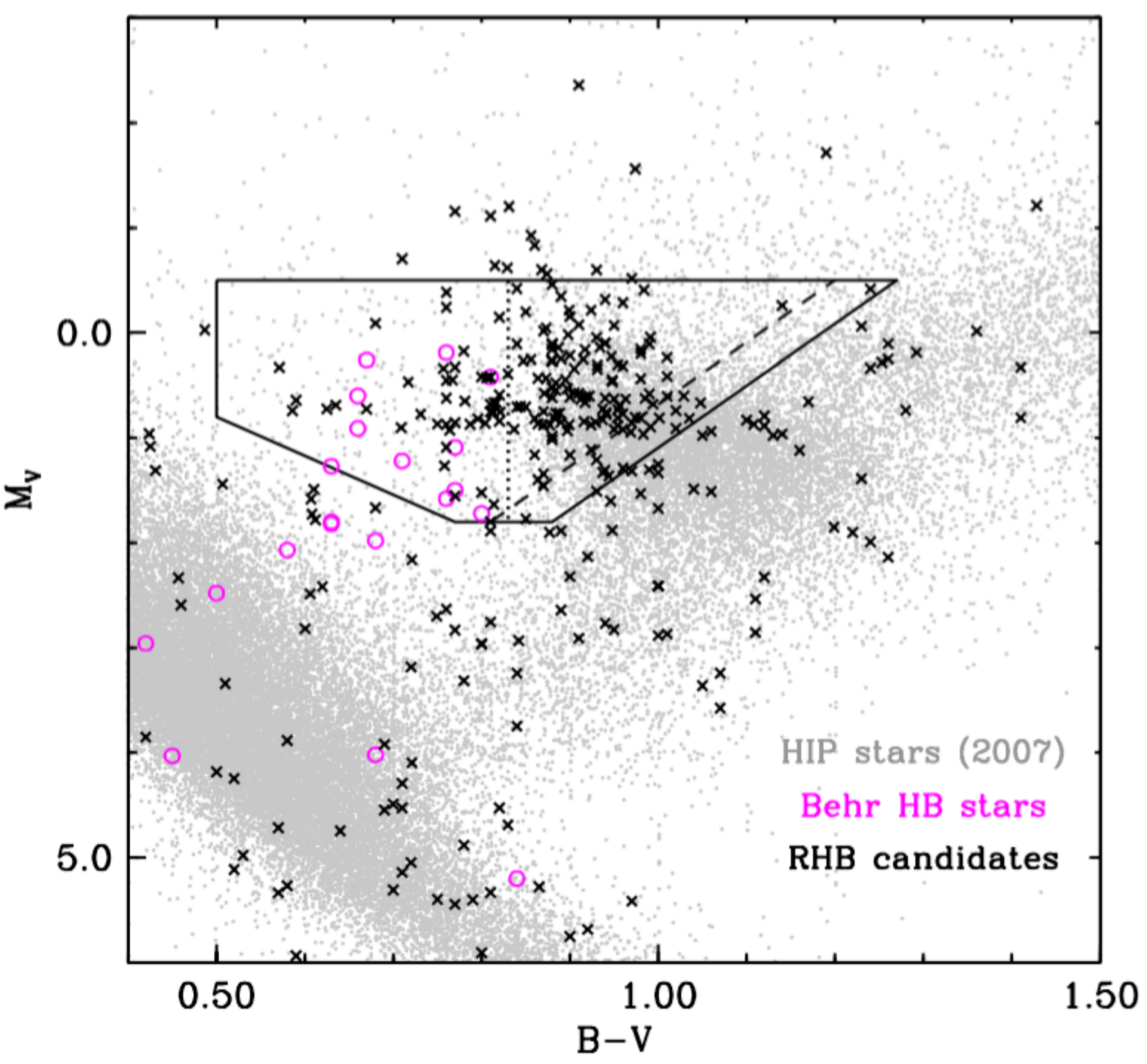}                                          
\begin{center}
\end{center}
\caption{\label{colormag}                                       
\footnotesize                                                 
    A color-magnitude diagram for the program stars.
    This figure is adapted from Figure~1 of \cite{kaempf05}.
    The solid, and broken lines that \citeauthor{kaempf05} used to define
    the RHB domain are reproduced in the present figure; see the text
    for a detailed description.
    The black crosses are from Table-\ref{tab-basic} data, the gray points
    are from the revised Hipparcos catalog \citep{vanleeuwen07}, and
    the magenta circles are from a targeted HB spectroscopic study by \cite{behr03b}.
}                                                             
\end{figure}

\begin{figure}                                                
\epsscale{1.00}                                              
\plotone{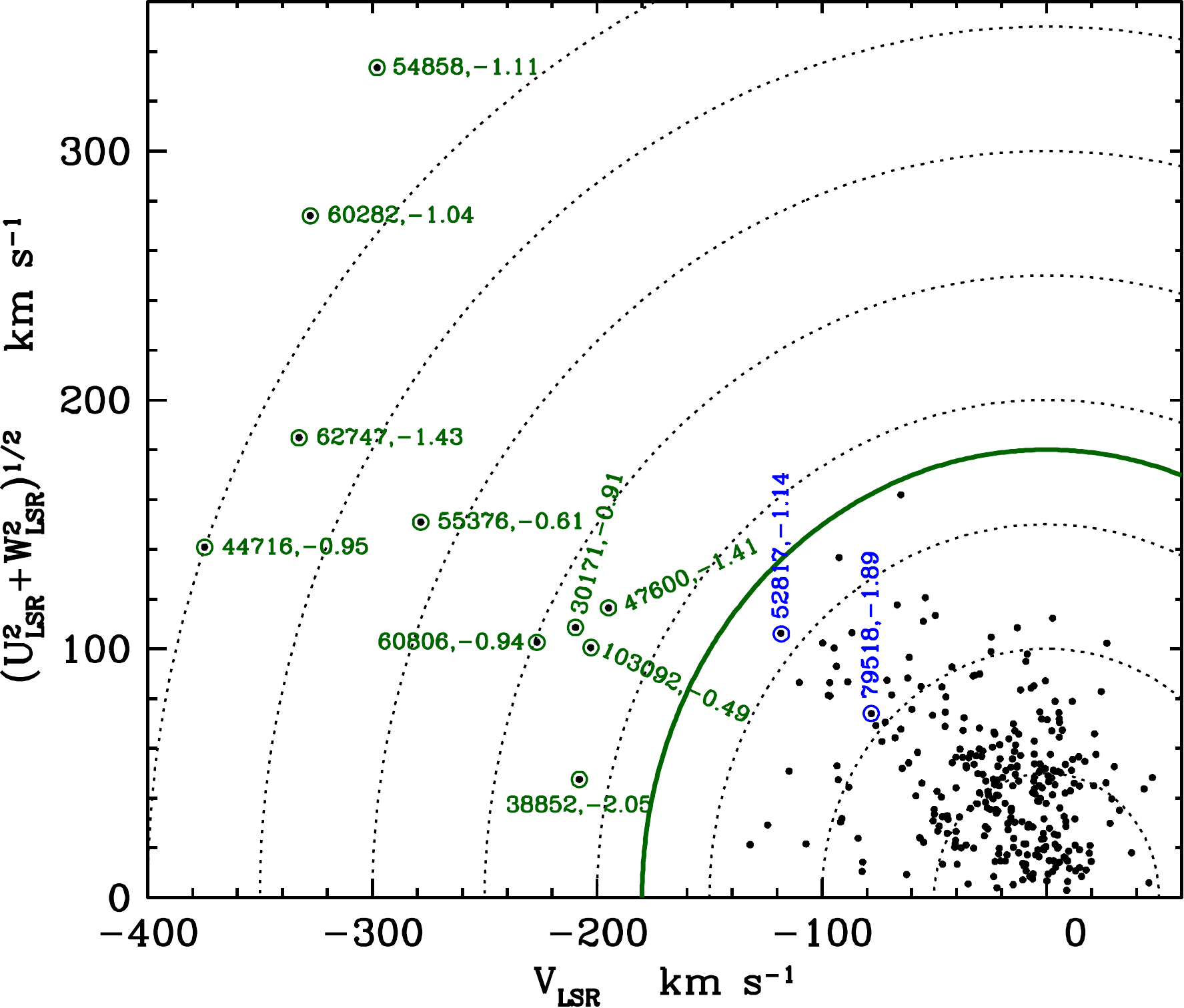}                                     
\begin{center}                                                
\end{center}                                                  
\caption{\label{toomre2}                                   
\footnotesize                                                 
    A Toomre diagram for all program stars.
    The dotted curves are placed at $V_{\rm{tot}}$ intervals of 50~\kmsec.
    The solid green curve is placed at $V_{\rm{tot}}$~=180~\kmsec, the approximate
    disk/halo transition velocity.
    Stars with $V_{\rm{tot}}$~$>$~180~\kmsec\ are labeled and two stars with
    smaller $V_{\rm{tot}}$ but [Fe/H]~$<$~$-$1 are labeled in blue (thick 
    disk members) with their Hipparcos names and metallicities from 
    Table~\ref{tab-models}.
}                                                             
\end{figure}

We selected the RHB candidates mainly using color 
temperature, luminosity, and surface gravity estimates;
no kinematical criteria were applied.
This selection was done several years before Gaia's first data release. 
Therefore our RHB absolute magnitude values included some rough 
estimates, especially for the stars without Hipparcos parallax information. 
With new Gaia data we were able to obtain more accurate parallaxes for more
program stars.
These showed general consistency with Hipparcos values for 
${\pi_{\rm Hip}}$~$>$~0.03~mas (Figure~\ref{paracomp}).
Out of 340 program stars, 309 have Gaia and/or Hipparcos values, making
it possible to construct a reliable CMD for more than 90\% of the RHB 
candidates (Figure~\ref{colormag}).
With the \cite{kaempf05} definition of the RHB CMD domain we were able
to separate the RHBs from
the subgiants and main sequence stars in our sample.

Spectra of high-resolution ($R$~$\simeq$~60,000) and
high $S/N$ (usually $\ge$~150) were obtained with the echelle 
spectrographs of the McDonald Observatory 2.7m HJS and 9.2m HET telescopes.
We also reanalyzed the spectra of 45 RHB stars studied by ASF12 in order 
to ensure internally consistent results among the whole RHB sample. 
From these spectra we determined the atmospheric parameters 
(\teff, \logg, \vmicro, [Fe/H]) and [X/Fe] abundance ratios for 
species \species{Fe}{i}, \species{Fe}{ii}, \species{Ti}{i}, \species{Ti}{ii}, 
\species{Cr}{i}, \species{Cr}{ii}, \species{Si}{i}, \species{Ca}{i}, 
and \species{Ni}{i}.
These analyses yielded 250 stars with 4500~$<$~\teff~$<$~5500 and 
\logg~$<$~3.5 (Figure~\ref{tefflogg}). 
Taking into account the estimated temperature and \logg\ limits for RHB stars
suggested by ASF12 we constrained the region of \textit{true RHB candidates} 
(tRHBc) in both the ($B-V$)~-~$M_V$ and \teff~-~\logg\ planes. 
This confinement yielded in the end about 150 tRHBc.

\begin{figure}                                                
\epsscale{1.00}                                              
\plotone{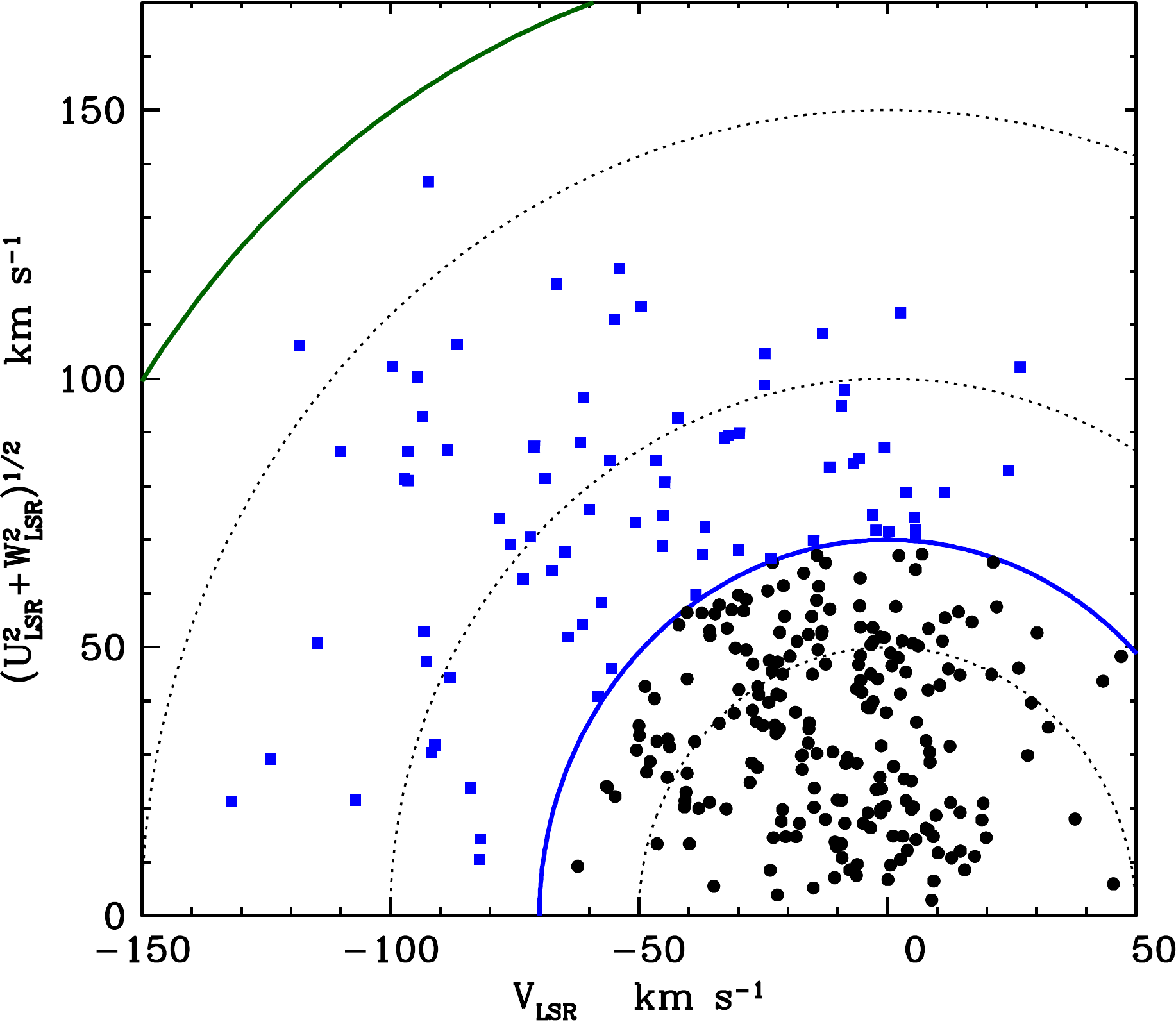}                                   
\begin{center}                                                
\end{center}                                                  
\caption{\label{toomre1}                                  
\footnotesize                                                 
    A Toomre diagram for only program stars with $V_{\rm{tot}}$~$<$~180~\kmsec.
    The dotted curves are lines of constant total velocities ($V_{\rm{tot}}$) with 
    intervals of 50~\kmsec.
    The solid green line is as in Figure~\ref{toomre2}.
    The solid blue line denotes the estimated approximate upper velocity 
    limit ($V_{\rm{tot}}$ = 70~\kmsec) for kinematically-defined thin disk 
    stars (shown with black dots). 
    The thick disk stars are shown with blue squares.
}                                                             
\end{figure}

\begin{figure}
\epsscale{1.00}
\plotone{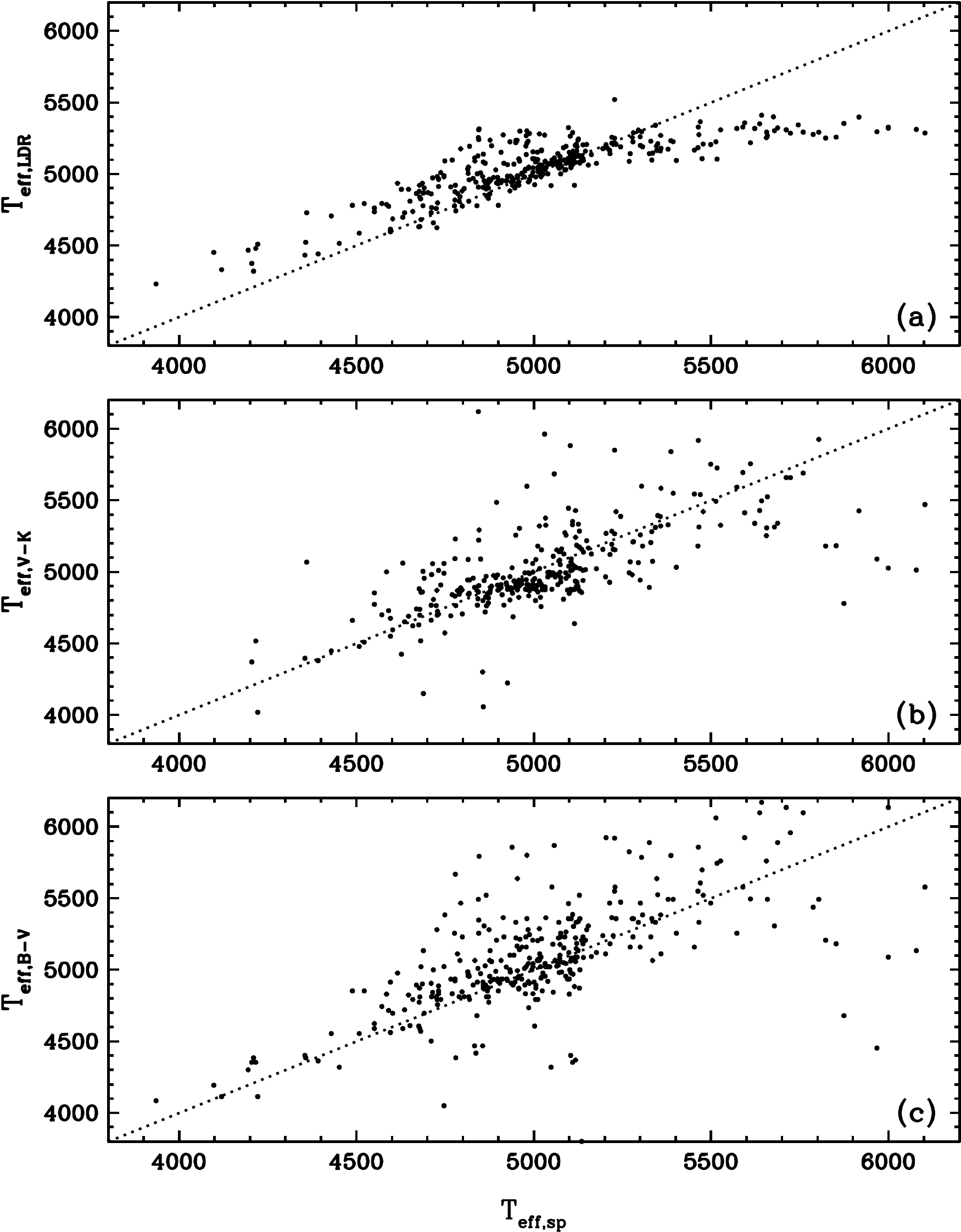}
\begin{center}
\end{center}
\caption{\label{teff3guess}
\footnotesize
    Comparison of final \teff$_{spec}$ values derived from spectroscopic 
    analyses with three input \teff\ indicators.
}
\end{figure}

\begin{figure}                                                  
\epsscale{1.00}                                                
\plotone{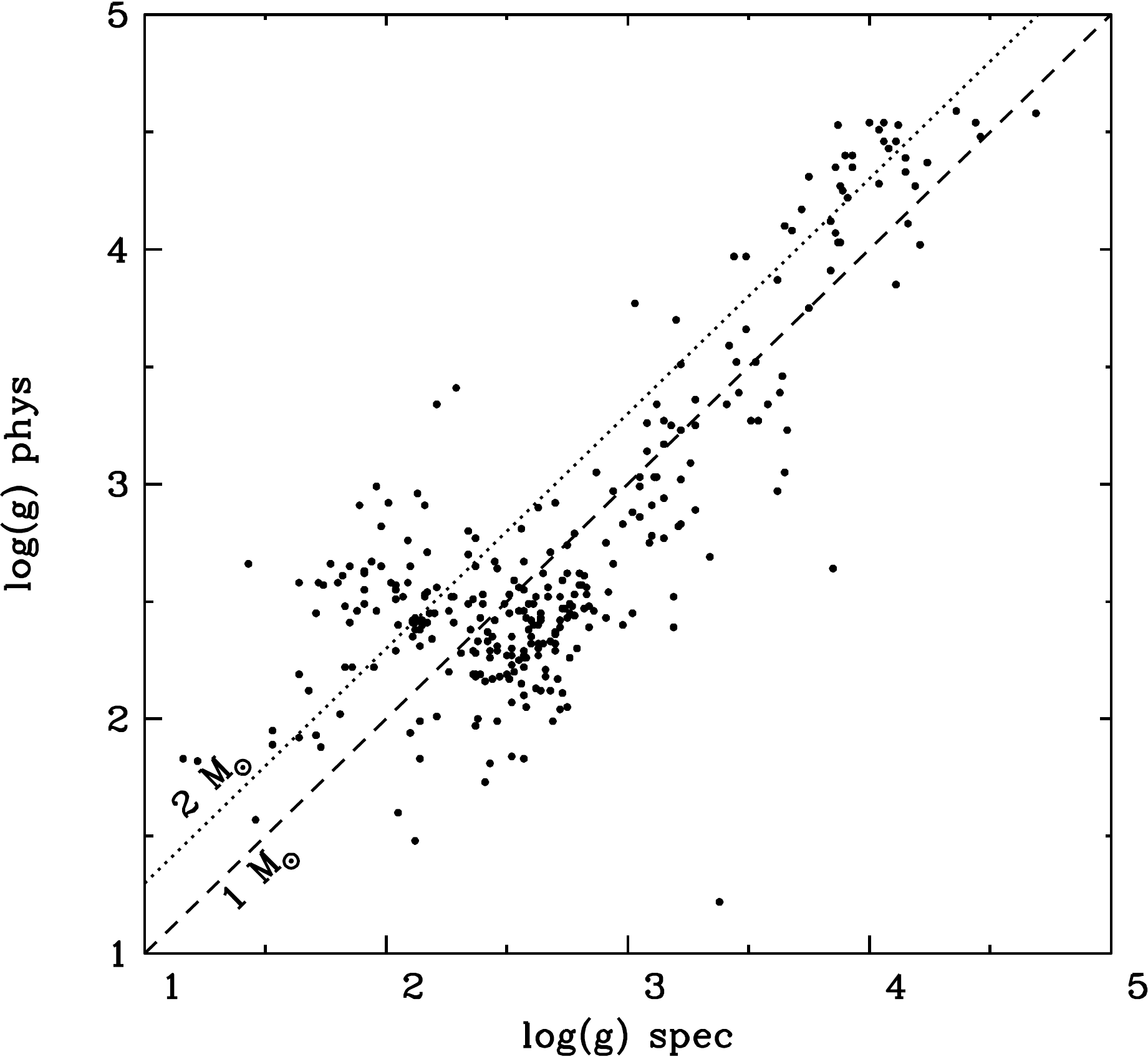}                                       
\begin{center}                                                  
\end{center}                                                    
\caption{\label{gphysgspec}                                     
\footnotesize                                                   
    Comparison of final \logg$_{spec}$ values derived from spectroscopic 
    analyses with the ``physical'' ones based on photometric and parallax data,
    and assumed solar masses, M~$\equiv$~ 1~M$_\odot$.
    Equality of the \logg\ values is indicated with the dashed line.
    The change in \logg$_{spec}$ that would occur if we had assumed 
    M~$\equiv$~ 2~M$_\odot$ is indicated with the dotted line.
}                                                               
\end{figure}

\begin{figure}
\epsscale{1.00}
\plotone{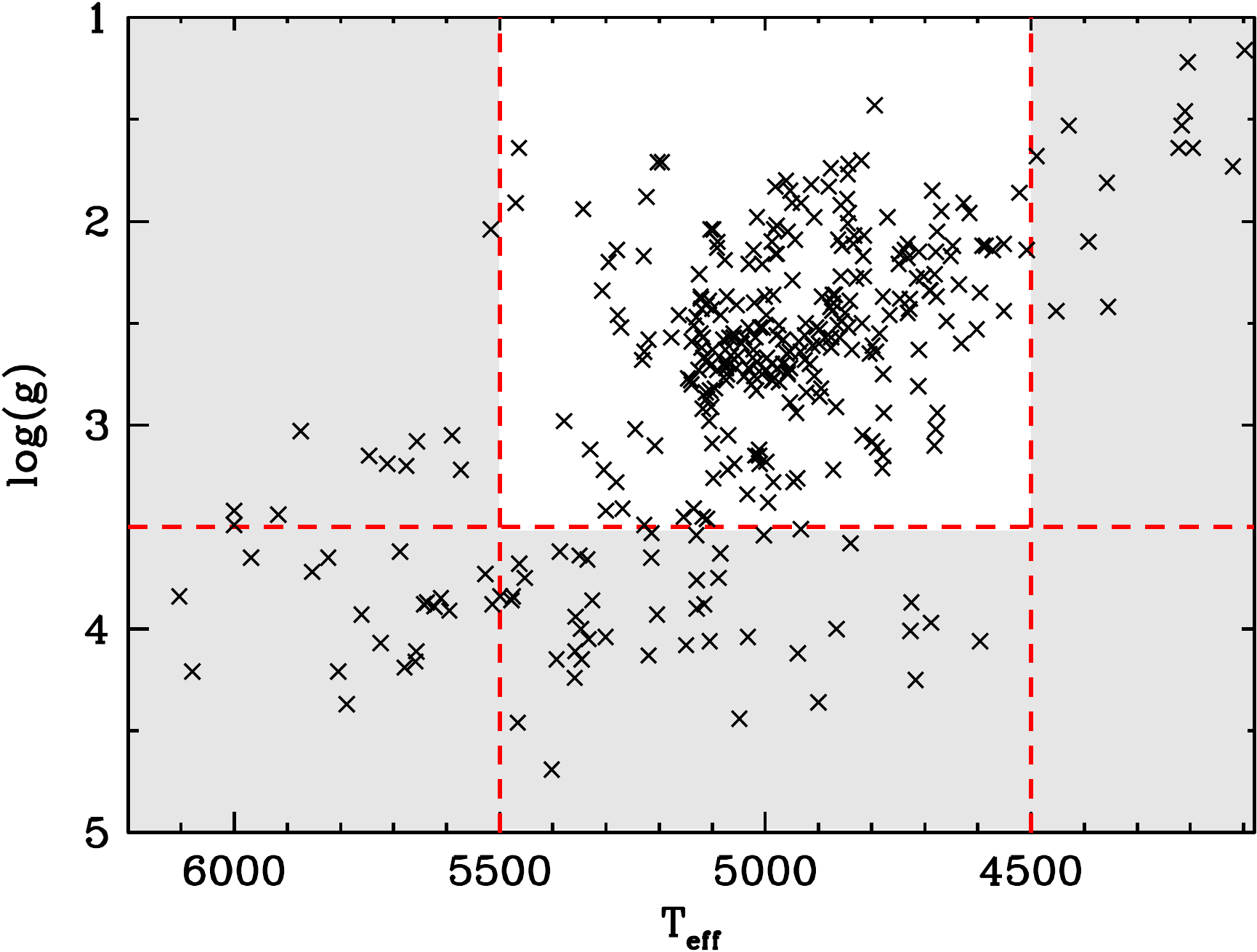}
\begin{center}
\end{center}
\caption{\label{tefflogg}
\footnotesize
    A \teff$-$ \logg\ HR diagram for the program stars.
    The data are taken from Table~\ref{tab-models}.
    Gray-shaded areas of the plot, those with \teff~$>$~5500~K or 
    \logg~$>$~3.5 are those regions with stars unlikely to be true
    RHB stars, and thus are of less interest in this study;
    see text for further discussion.
}
\end{figure}

\begin{figure}
\epsscale{1.00}
\plotone{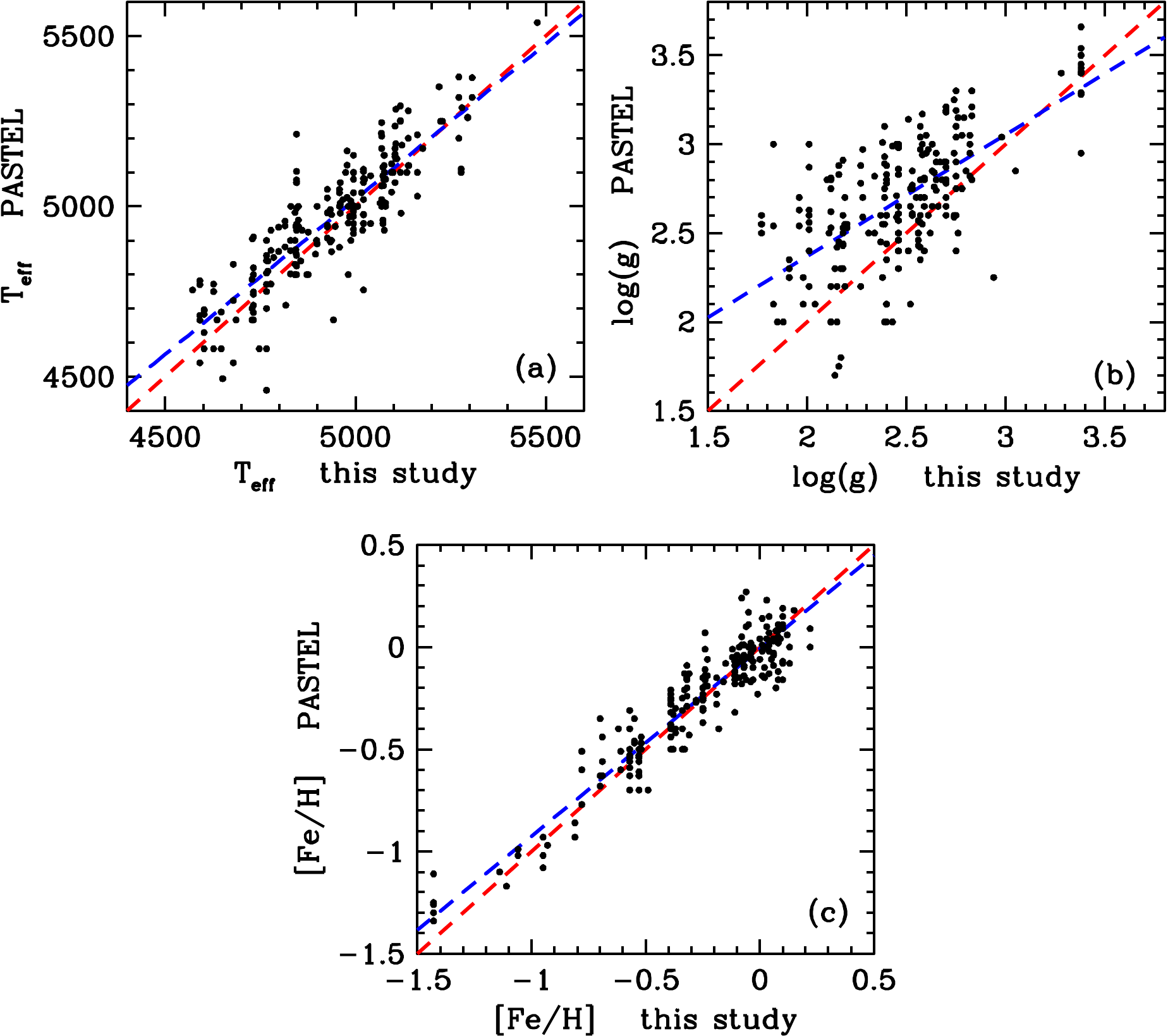}
\begin{center}
\end{center}
\caption{\label{pastelcomp}
\footnotesize
    Comparisons of model parameters derived in this paper with ones in
    the PASTEL catalog \citep{soubiran16}.
    In each panel the red solid line represents equality of the quantities,
    and the blue dashed line represents the mean offset from equality.
}
\end{figure}

\begin{figure}
\epsscale{1.00}
\plotone{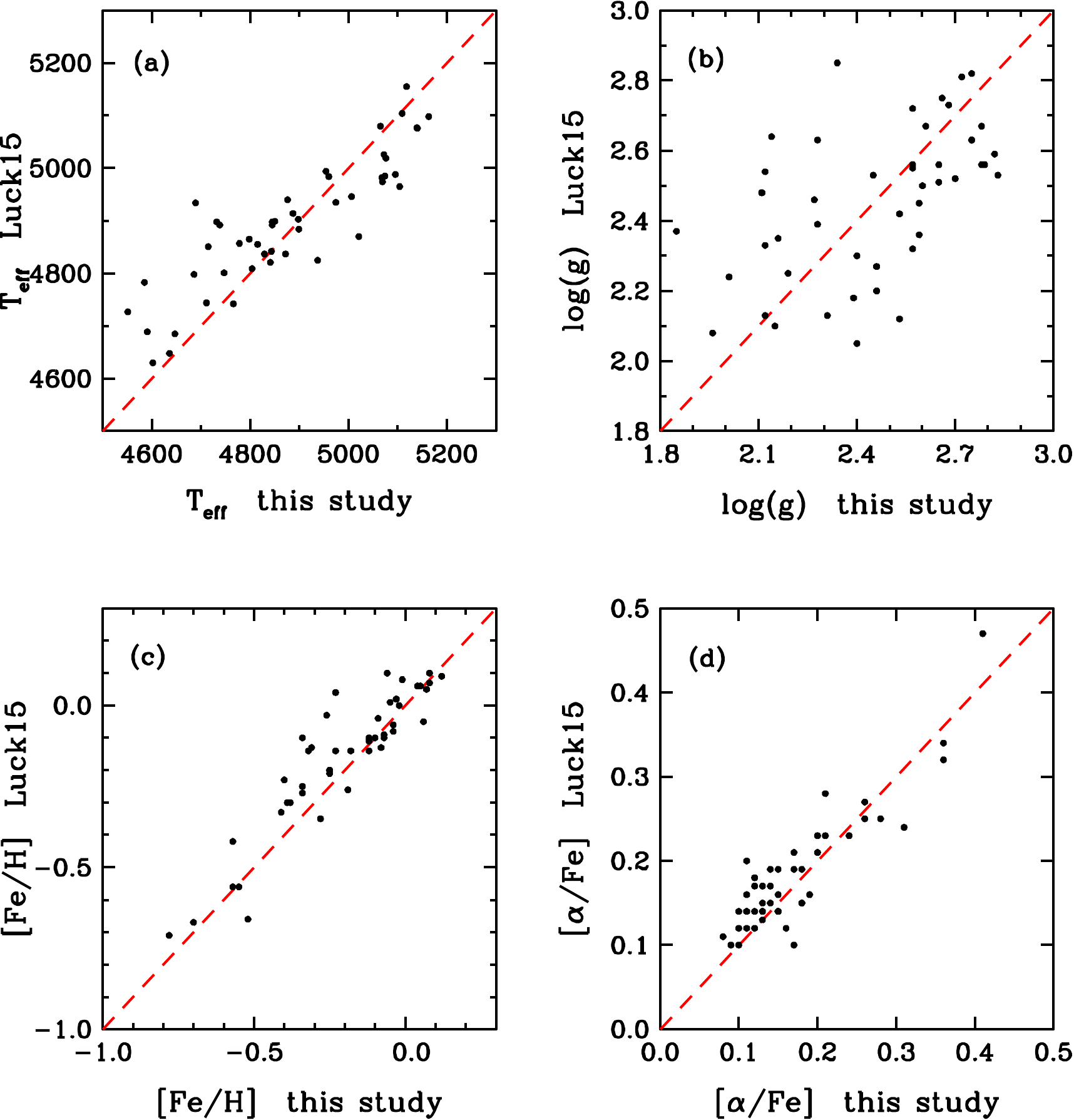}
\begin{center}
\end{center}
\caption{\label{luckcomp}
\footnotesize
    Comparisons of model parameters derived in this paper and [$\alpha$/Fe]
    abundance ratios with ones in \cite{luck15}.
    In each panel the red solid line represents equality of the quantities.
}
\end{figure}

\begin{figure}
\epsscale{1.00}
\plotone{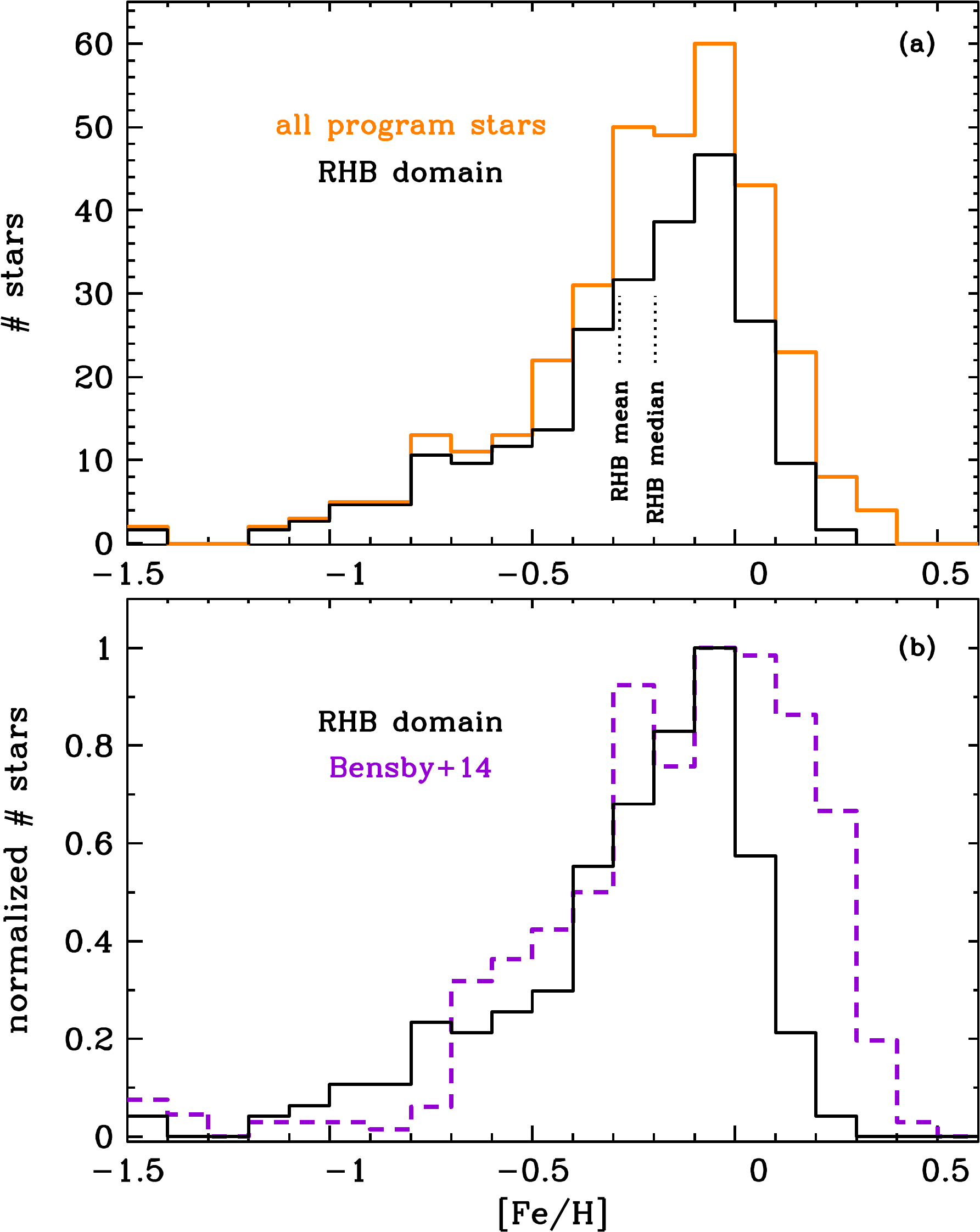}
\begin{center}
\end{center}
\caption{\label{fehisto}
\footnotesize
    Histograms of the [Fe/H] metallicity distribution of our program
    stars.
    The orange ``all stars'' line is for our total stellar sample.
    The black ``RHB domain'' line is for the subset of stars with
    a reasonable chance to be RHB stars, as defined in 
    Figure~\ref{tefflogg}. Dashed (purple) line represents the sample from
    \cite{bensby14}. 
}
\end{figure}

\begin{figure}
\epsscale{1.00}
\plotone{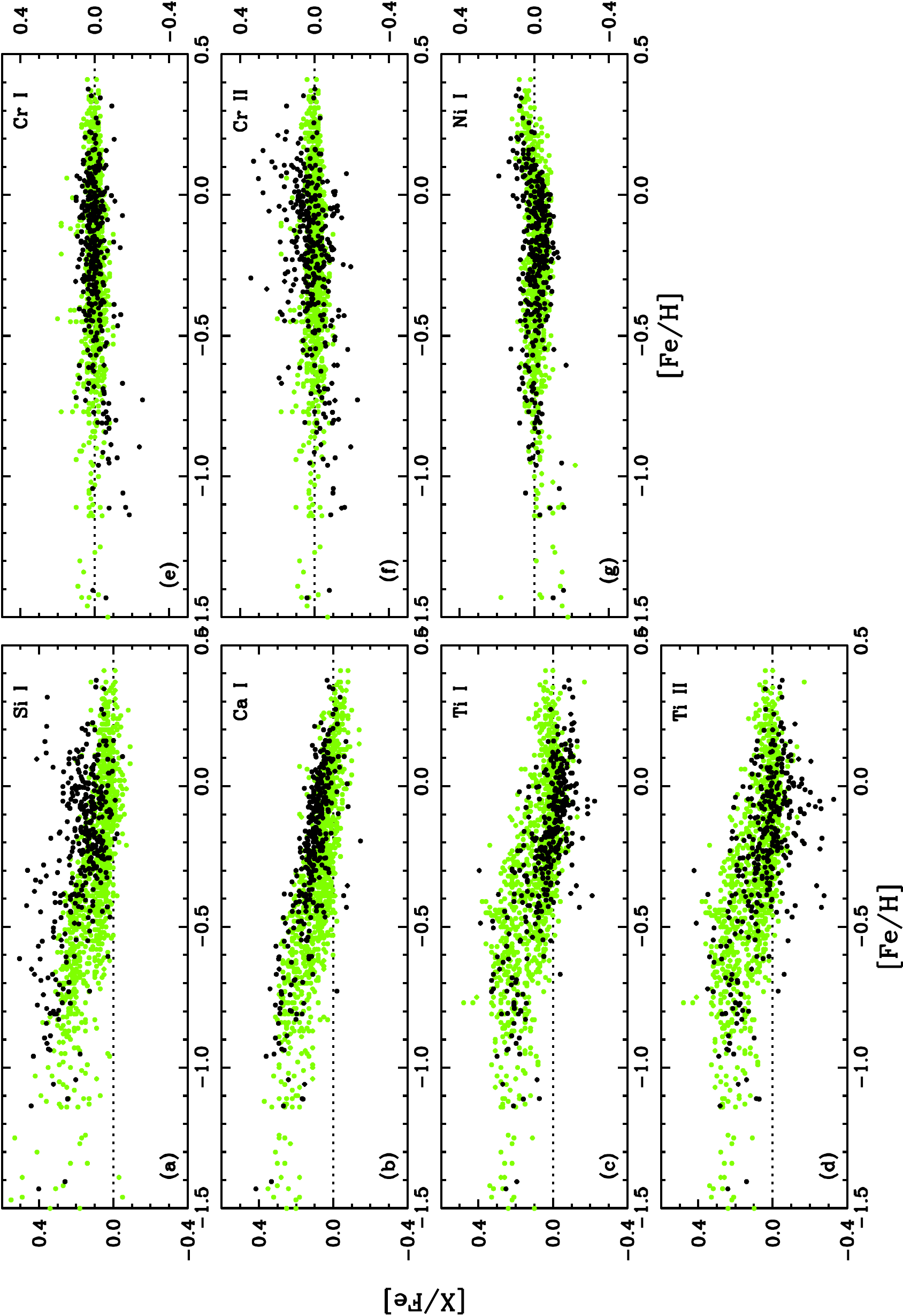}
\begin{center}
\end{center}
\caption{\label{allelements}
\footnotesize
    Correlations of [Fe/H] metallicity with relative abundance ratios 
    [X/Fe] of elements studied in this paper.
    Black points are our results, and chartreuse points are taken from
    surveys of Galactic disk main sequence and early subgiants studied by
    \cite{reddy03,reddy06} and \cite{bensby14}.
    The dotted lines represent the solar abundance ratios, [X/Fe]~=~0.
}
\end{figure}

\begin{figure}
\epsscale{1.00}
\plotone{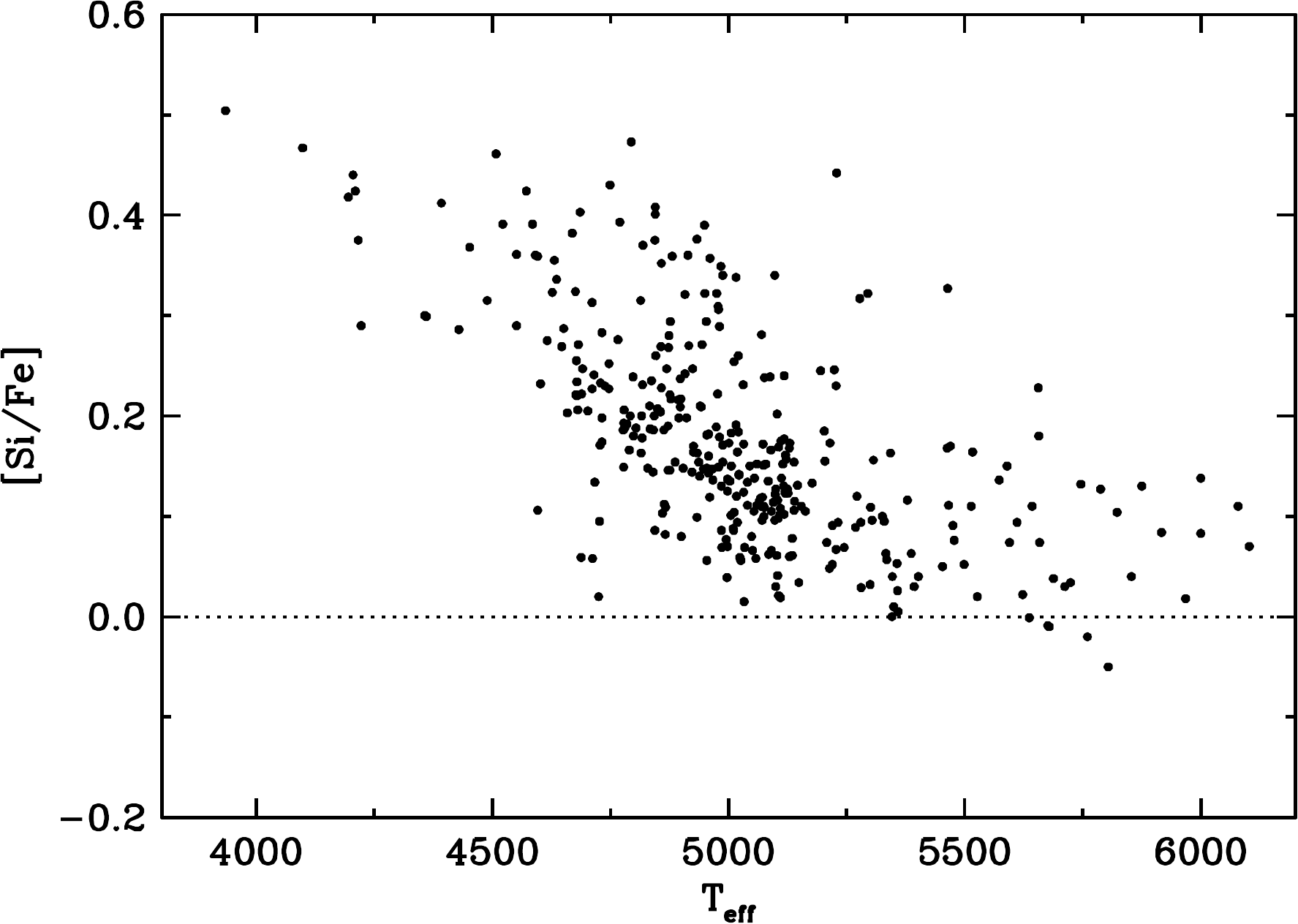}
\begin{center}
\end{center}
\caption{\label{Teff_Si}
\footnotesize
      [Si/Fe] ratios plotted as a function of \teff.
}
\end{figure}

We calculated space velocity components ($U_{\rm{LSR}}$, 
$V_{\rm{LSR}}$, $W_{\rm{LSR}}$ ) for 305 stars (see \S\ref{motions}) 
using the distances and 
proper motions from Gaia or Hipparcos, and RVs derived from our spectra 
(\S\ref{obsred}).
From the Toomre diagram (Figure~\ref{toomre1}), we have concluded that
our sample is made up of 215 thin disk, 80 thick disk and 
10 halo stars. About 80\% of the tRHBc appears in kinematical 
thin disk, while the rest are thick disk members.

\begin{figure}
\epsscale{1.00}
\plotone{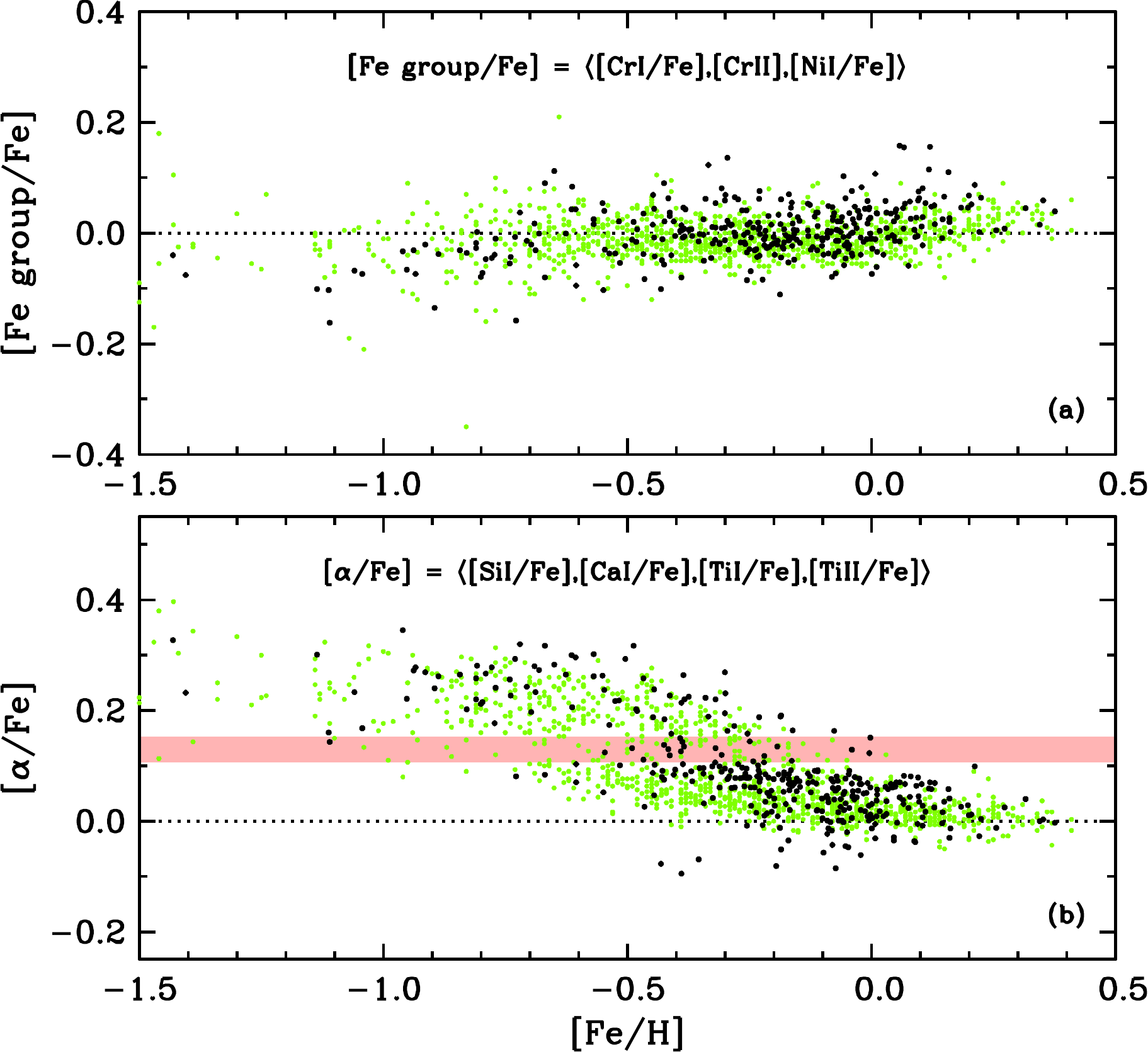}
\begin{center}
\end{center}
\caption{\label{means}
\footnotesize
    Correlations of [Fe/H] with [X/Fe] for elements of the Fe group (panel a) 
    and the $\alpha$'s (panel b) studied in this paper.
    The species participating in the means are labeled in the figure.
    The point types and black dotted lines are as in Figure~\ref{allelements}.
    In panel (b) a pale red line has been placed at [$\alpha$/Fe]~=~$+$0.13,
    the approximate abundance-based dividing line for the $\alpha$-enhanced 
    group.
}
\end{figure}

\begin{figure}
\epsscale{1.00}
\plotone{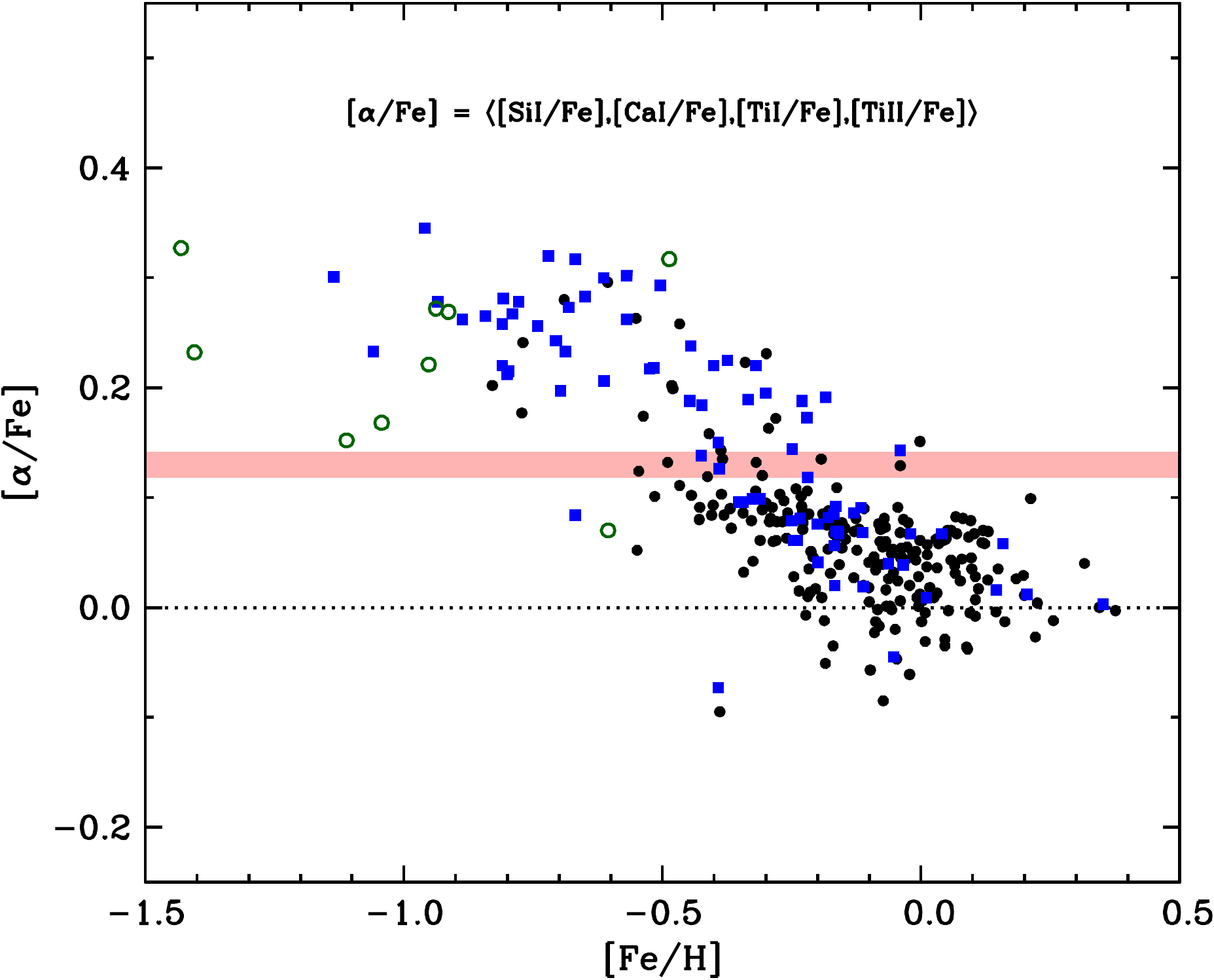}
\begin{center}
\end{center}
\caption{\label{mean_kin}
\footnotesize
    Figure~\ref{means} replotted with kinematic group 
    color-coding.  
    The symbols and colors are the same as given in Figure~\ref{toomre1}. 
    Halo stars are denoted with empty green circles, as in 
    Figure~\ref{toomre2}.  
}
\end{figure}

\begin{figure}                                                
\epsscale{1.0}                                              
\plotone{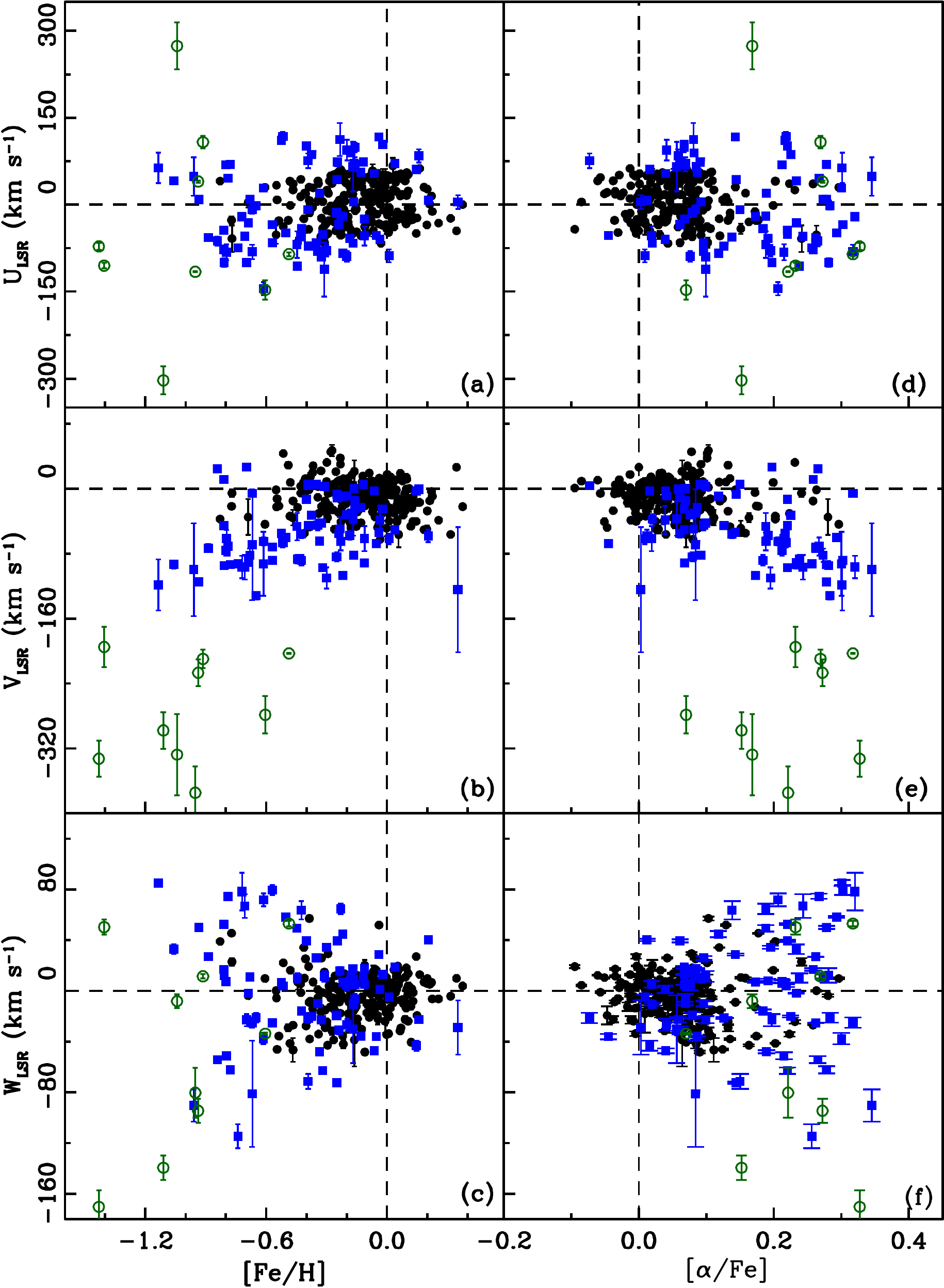}                                    
\begin{center}                                                
\end{center}                                                  
\caption{\label{FeH_alpha_kin}                                  
\footnotesize                                                 
    Correlation of [Fe/H] with (a) $U_{LSR}$, 
    (b) $V_{LSR}$, (c) $W_{LSR}$, and correlation of [$\alpha$/Fe] 
    abundance ratios with (d) $U_{LSR}$, (e) $V_{LSR}$, and (f) $W_{LSR}$.
    The symbols and colors are the same as in Figure~\ref{mean_kin}. 
}                                                             
\end{figure}

Our kinematically unbiased sample 
has representation over a wide metallicity range,
$-$2.0~$\lesssim$~[Fe/H]~$\lesssim$~$+$0.4 (Table \ref{tab-models}), but
the majority of the stars have near-solar metallicities 
($\langle$[Fe/H]$\rangle$~$\sim$~$-$0.2, $\sigma$~$\sim$~0.3).
We have only a few halo stars with [Fe/H]~$<$~$-$1.0. 
General metallicity and [$\alpha$/Fe] ratio distributions (Figure~\ref{means}b) 
of our disk stars are broadly similar to other studies that investigate larger,
more statistically robust samples, such as, \cite{liu12}, \cite{bensby14}, 
\cite{anders14}, \cite{luck15} and \cite{hayden15}.
Even with our relatively small sample size we can see the bimodality in 
[$\alpha$/Fe] at low metallicities and its extension towards 
[Fe/H]~$\simeq$~$-$0.6 discussed by, \eg, \citealt{nidever14,bensby14}, 
and references cited in those studies.
The ``knee" at [Fe/H]~$\simeq$~$-$0.5 (\citealt{bensby03,reddy06,bensby17}) 
also clearly appears in our sample. 
On the other hand, the Fe-group elements Cr and Ni have less variation with 
metallicity (Figure~\ref{means}a); their abundance ratios are solar in 
all [Fe/H] regimes of this study, in agreement with, \eg, 
\citealt{reddy06,bensby14}.
Additionally, the chemically thin- and thick disk stars are on average
kinematically separated:  the high-$\alpha$ stars have larger space velocities,
as seen in Figure~\ref{mean_kin}.

\begin{figure}                                                
\epsscale{1.00}                                              
\plotone{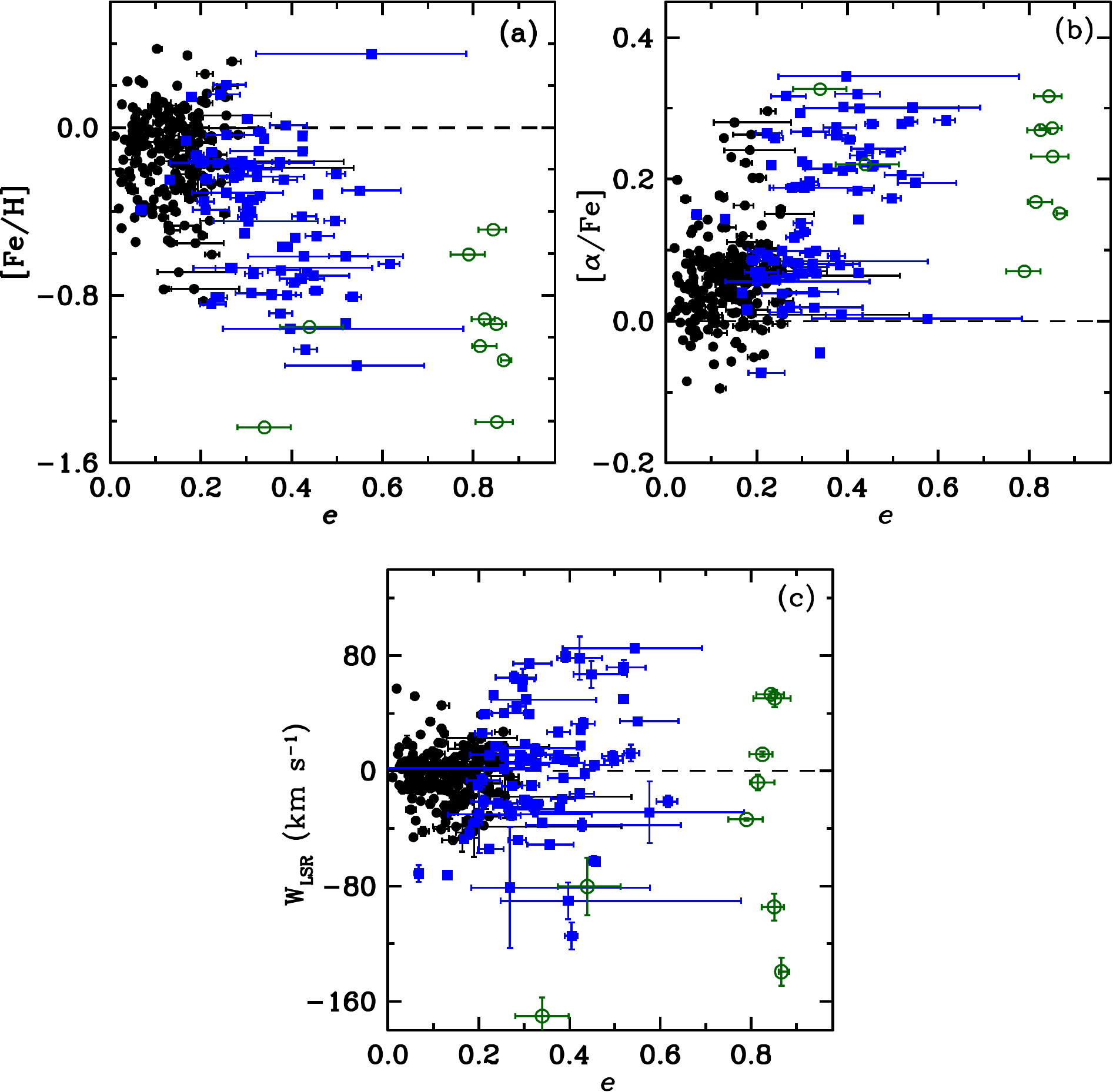}                                    
\begin{center}                                                
\end{center}                                                  
\caption{\label{FeHalpW_ecc}                                  
\footnotesize                                                 
    Galactic orbital eccentricity, $e$, versus (a) [Fe/H], (b) [$\alpha$/Fe], 
    and (c) $W_{LSR}$. The symbols and colors are the same as in 
    Figure~\ref{mean_kin}.
}                                                             
\end{figure}

\begin{figure}                                                
\epsscale{1.00}                                              
\plotone{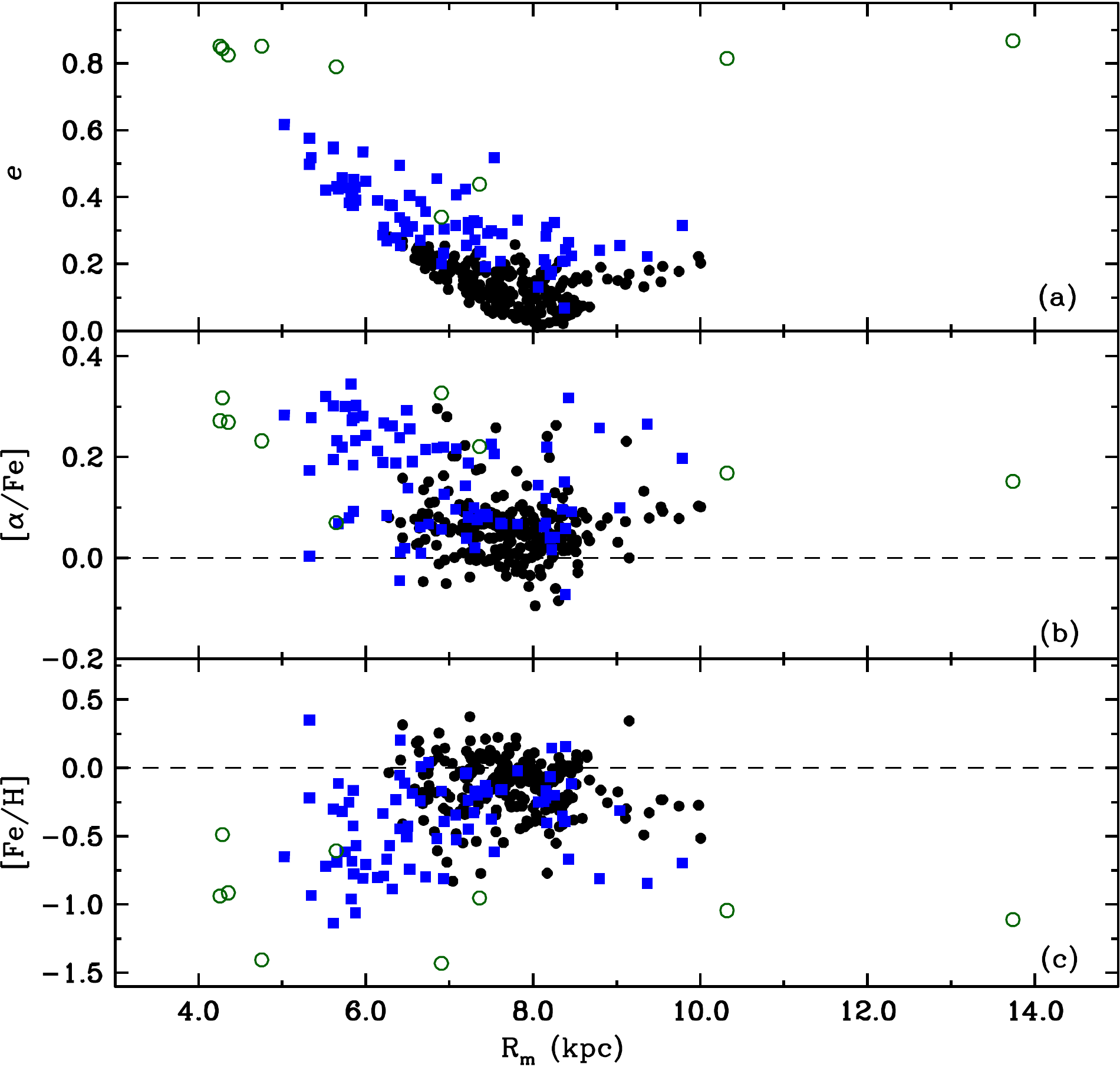}                                    
\begin{center}                                                
\end{center}                                                  
\caption{\label{Rm_e_alp_Fe}                                  
\footnotesize                                                 
    $e$, [$\alpha$/Fe] and Fe/H] plotted as a function of R$_{\rm{m}}$ 
    (mean distance to the Galactic center). The symbols and colors are 
    the same as in Figure~\ref{mean_kin}.
}                                                             
\end{figure}

Investigation of $\alpha$ abundances as functions of mean Galactocentric
distances (Figure~\ref{Rm_e_alp_Fe}) indicated that the $\alpha$-rich group 
resides mostly at R$_{\rm{m}}$~$<$~7 kpc, while $\alpha$-solar stars 
mostly occupy the solar radius and some at R$_{\rm{m}}$~$>$~9 kpc.
The $\alpha$-solar group ([Fe/H]~$>$~$-$0.3, 
[$\alpha$/Fe]~$\lesssim$~$+$0.13) is mostly composed of kinematically 
thin disk stars.
It also includes a number of thick disk stars with thin disk chemical 
compositions. 
The common feature of these stars that they have orbital eccentricities
$e$~$<$~0.3 (Figure~\ref{FeHalpW_ecc}) and they are mostly within the solar 
circle (7 kpc~$<$~R$_{\rm{m}}$~$<$~9 kpc, Figure~\ref{Rm_e_alp_Fe}),
compared to their high eccentric counterparts.
These $\alpha$-solar stars have $|W_{\rm{LSR}}|$ $\lesssim$~40 \kmsec, 
which indicates that they are kinematically located at smaller vertical 
scale heights and contribute mostly to the disk-like rotation. 
Regardless of their kinematic definitions, the stars around the solar 
neighbourhood may share a similar formation history that involves a 
metallicity enriched ISM via SNe Ia. Alternatively, as discussed in many studies 
(\eg, \citealt{sell02,roskar08,hay08,schon09,loebman11,
lee11,boeche13,hayden15,daniel15,kordo15}, and references therein),
old thick disk stars that were originally formed from enriched gas far
 from the solar neighborhood towards the Galactic center may 
appear in the solar radius due to radial migration of stars. 
Having a significant amount of kinematically thick disk stars at 
$\alpha$-solar group may be also explained via radial migration of 
group of stars with originally thin disk kinematics now appearing at 
locations vertically more distant to the Galactic plane.

\begin{figure}                                                
\epsscale{1.00}                                              
\plotone{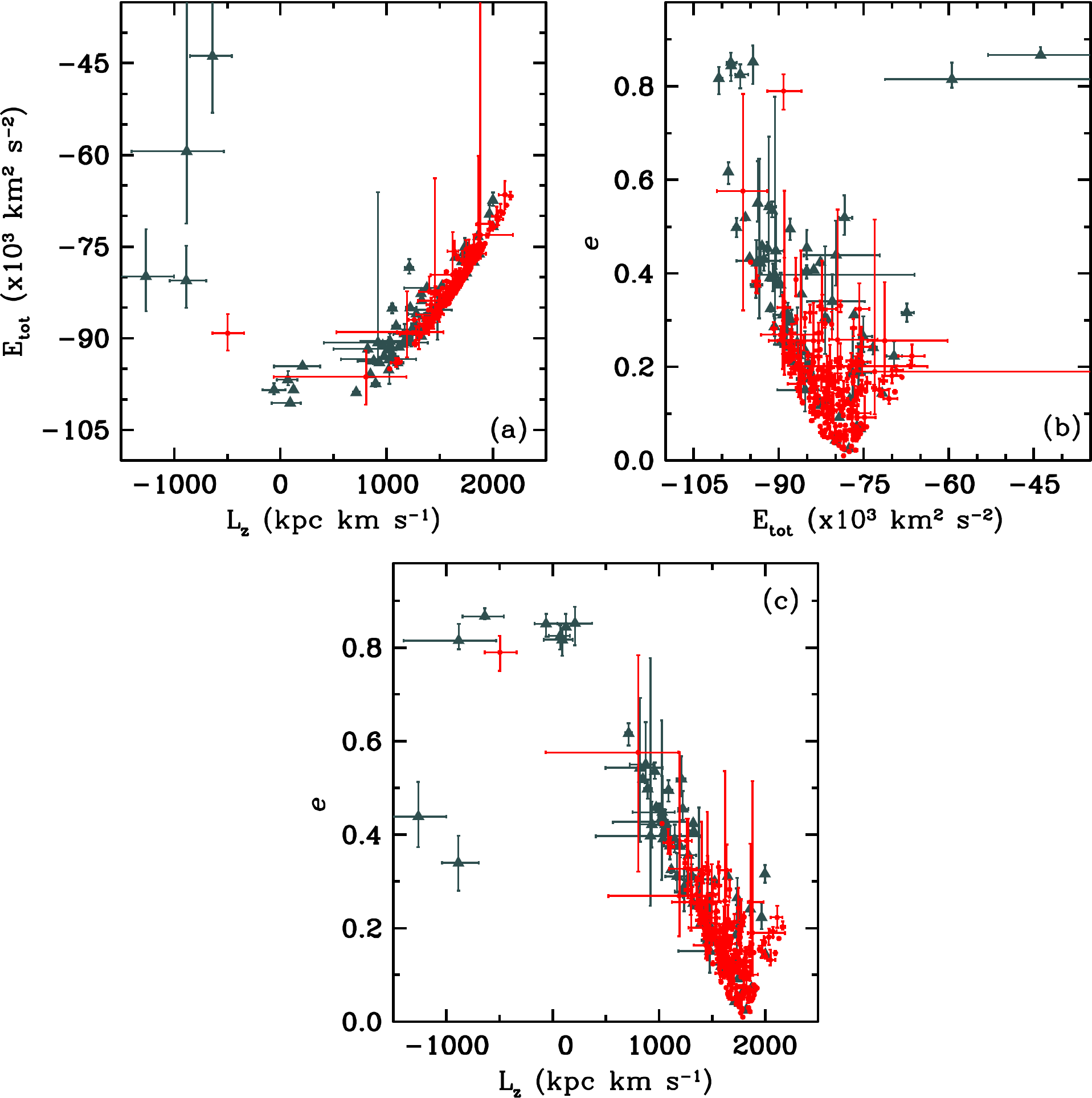}                                    
\begin{center}                                                
\end{center}                                                  
\caption{\label{alpharp_Evt_Lz}                                  
\footnotesize                                                 
    (a) E$_{\rm{tot}}$ $-$ L$_{\rm{z}}$. (b) $e$ $-$ E$_{\rm{tot}}$, and 
    (c) $e$ $-$ L$_{\rm{z}}$ relations for the whole sample, including two 
    low-metallicity stars with [Fe/H] $<$ $-$1.5. 
    Dark grey triangles denote $\alpha$-rich ([$\alpha$/Fe]~$>$~$+$0.13), red 
    dots denote $\alpha$-solar ([$\alpha$/Fe]~$\lesssim$~$+$0.13) stars. 
}                                                             
\end{figure}

The E$_{\rm{tot}}$ - L$_{\rm{z}}$ plot in 
Figure~\ref{alpharp_Evt_Lz}a shows that a few $\alpha$-rich stars 
and one $\alpha$-solar star are in retrograde orbits. 
There are five $\alpha$-rich stars that appear clumped in the phase space 
(\ie, with near zero L$_{\rm{z}}$, low orbital energy and high eccentricity). 
Their Galactocentric radii R$_{\rm{m}}$~$\sim$~4.3 kpc 
indicate that they spend most of their time in the 
inner regions of the Galaxy.
Most of our sample is clustered at high L$_{\rm{z}}$ values, which increase 
with E$_{\rm{tot}}$ in disk-like rotations.
$\alpha$-solar stars have circular/near-circular orbits ($e$~$<$~0.3)
(Figure~\ref{alpharp_Evt_Lz}a), while $\alpha$-rich stars have more eccentric 
orbits with less angular momenta implying less disk-like rotation. 
The $\alpha$-rich stars that have similar energies, angular momenta 
and eccentricities to those of $\alpha$-solar 
stars have mostly disk-like 
rotation and correspond to the high-metallicity tail of the kinematically 
thick disk stars.

The chemo-kinematical analysis of more than 300 RHB 
candidates allowed us to define the tRHBc in our sample. 
We identified about 150 tRHBc according to their HR diagram locations 
based on both ($B-V$)~-~$M_V$ and \teff~-~\logg\ criteria. 
The majority of them are kinematically thin disk members, while some has 
the kinematical thick disk membership. 
The second paper \citep{afsar18} of this series will focus on the detailed chemical 
abundance analysis of these tRHBc. 
We will investigate CNO abundances and \carbiso\ ratios along with many 
other elements such as $\alpha$ element Mg to complement Ca and Si reported
in this paper; Fe-group elements: Sc, V, Mn, Co, Cu, Zn; 
and neutron-capture elements: Y, La, Nd, Eu.
With the CNO abundances and \carbiso\ ratios we will be able to 
investigate the evolutionary stages in detail and have more robust 
analysis to identify the tRHBc in our sample, and deepen the 
chemo-kinematic analysis we obtained in this study.

\acknowledgments

We thank Tim Beers, Earle Luck, and our referee for 
helpful discussions and comments on the manuscript.
Our work has been supported by The Scientific and Technological
Research Council of Turkey (T\"{U}B\.{I}TAK, project No. 112T929), by the US
National Science Foundation grants AST~1211585 and AST~1616040, and by the
University of Texas Rex G. Baker, Jr. Centennial Research Endowment.
This research has made use of the SIMBAD database, operated at CDS, 
Strasbourg, France. 
This work has made use of data from the European Space 
Agency (ESA) mission {\it Gaia} (\url{https://www.cosmos.esa.int/gaia}), processed by
the {\it Gaia} Data Processing and Analysis Consortium (DPAC,
\url{https://www.cosmos.esa.int/web/gaia/dpac/consortium}). Funding
for the DPAC has been provided by national institutions, in particular
the institutions participating in the {\it Gaia} Multilateral Agreement.

\software{IRAF (\citealt{tody93} and references therein),
          MOOG \citep{sneden73},
          ATLAS \citep{kurucz11}}

\bibliographystyle{apj}
\bibliography{totbib} 


\newpage

\clearpage

\clearpage
\begin{center}
\begin{deluxetable}{lllllllcccl}
\tabletypesize{\footnotesize}
\tablewidth{0pt}
\tablecaption{Basic Program Star Data\label{tab-basic}}
\tablecolumns{11}
\tablehead{
\colhead{Star\tablenotemark{a}}         &
\colhead{RA$_{2000}$\tablenotemark{b}}  &
\colhead{Dec$_{2000}$\tablenotemark{b}} &
\colhead{$B$\tablenotemark{b}}          &
\colhead{$V$\tablenotemark{b}}          &
\colhead{$K$\tablenotemark{b}}          &
\colhead{$A_V$}                         &
\colhead{ref($A_V$)}                    &
\colhead{$E(B-V)$\tablenotemark{c}}     &
\colhead{spec\tablenotemark{d}}         &
\colhead{inst\tablenotemark{d}}        
}
\startdata
HIP 258	&	00 03 13.32	&	$+$17 33 08.16	&	7.69	&	6.57	&	4.36	&		&		&		&	K0	&	2.7m/Tull	\\
HIP 410	&	00 05 01.19	&	$+$27 40 29.22	&	7.41	&	6.48	&	4.27	&		&		&		&	G8III	&	2.7m/Tull	\\
HIP 476	&	00 05 41.96	&	$+$13 23 46.56	&	6.43	&	5.55	&	3.77	&		&		&		&	G5III	&	2.7m/Tull	\\
HIP 716	&	00 08 52.14	&	$+$25 27 46.74	&	7.22	&	6.25	&	4.20	&		&		&		&	K0III	&	2.7m/Tull	\\
HIP 1009	&	00 12 34.17	&	$+$44 42 25.90	&	7.64	&	6.52	&	4.51	&		&		&		&	K0	&	2.7m/Tull	\\
...
\enddata
\tablerefs{1 - \cite{bailerjones11}; 2 - \cite{jofre15}; 3 - \cite{straizys05};
           4 - \cite{liu10}; 5 - \cite{takeda08}; 6 - \cite{matrozis13};
           7 - \cite{goswami13}; 8 - \cite{wang11}; 9 - \cite{cruzalebes13};
           10 - \cite{sanchezblazquez06}; 11 - \cite{huber12}}
\tablenotetext{a}{When available, HIP \# preferably, then BD \#, then GSC \#}
\tablenotetext{b}{SIMBAD; spec = spectral type}
\tablenotetext{c}{$E(B-V)$~=~$A_V$/3.1}
\tablenotetext{d}{2.7m/Tull = McDonald 2.7m telescope and \cite{tull95} 
                  echelle spectrograph; HET/HRS = Hobby-Eberly Telescope and 
                  \cite{tull98} echelle spectrograph}
\vspace{1ex}
      \item (This table is available in its entirety in machine-readable form.)              
\end{deluxetable}
\end{center}

\begin{center}
\begin{deluxetable}{lccccrcrrrrrr}
\tabletypesize{\footnotesize}
\tablewidth{10pt}
\tablecaption{Parallaxes and Velocities \label{tab-motions}}
\tablecolumns{13}
\tablehead{
\colhead{Star}                          &
\colhead{$\pi_{\rm Hip}$}                         &
\colhead{err($\pi_{\rm Hip}$)}                 &
\colhead{$\pi_{\rm Gaia}$}                         &
\colhead{err($\pi_{\rm Gaia}$)}                 &
\colhead{$RV$}                         &
\colhead{$\sigma RV$}                  &
\colhead{$U_{LRS}$}                     &
\colhead{$\sigma U_{LRS}$}              &
\colhead{$V_{LRS}$}                     &
\colhead{$\sigma V_{LRS}$}              &
\colhead{$W_{LRS}$}                     &
\colhead{$\sigma W_{LRS}$}              \\
\colhead{}                              &
\colhead{mas}                           &
\colhead{mas}                           &
\colhead{mas}                          &
\colhead{mas}                          &
\colhead{\kmsec}                              &
\colhead{\kmsec}                              &
\colhead{\kmsec}                        &
\colhead{\kmsec}                        &
\colhead{\kmsec}                        &
\colhead{\kmsec}                        &
\colhead{\kmsec}                        &
\colhead{\kmsec}
}
\startdata
HIP 258	&	5.83	&	0.42	&	6.98	&	0.47	&	 $-$18.9	&	0.19	&	 $-$5.44	&	1.30	&	 $-$46.29	&	2.58	&	 $-$12.25	&	2.18	\\
HIP 410	&	5.95	&	0.42	&		&		&	14.60	&	0.14	&	 $-$49.46	&	3.91	&	 $-$15.87	&	2.30	&	 $-$17.34	&	1.17	\\
HIP 476	&	8.75	&	0.30	&		&		&	2.31	&	0.12	&	 $-$9.55	&	0.67	&	 $-$6.04	&	0.45	&	 $-$0.80	&	0.24	\\
HIP 716	&	7.22	&	0.32	&	7.50	&	0.57	&	15.05	&	0.17	&	 $-$65.43	&	5.38	&	2.34	&	1.08	&	14.73	&	1.25	\\
HIP 1009	&	7.10	&	0.63	&	7.47	&	0.53	&	 $-$35.55	&	0.17	&	 $-$23.54	&	3.45	&	 $-$56.28	&	2.22	&	 $-$4.16	&	1.58	\\
...
\enddata
\tablenotetext{a}{when available, using $\pi_{\rm Gaia}$, else $\pi_{\rm Hip}$.
                  }
\vspace{1ex}
      \item (This table is available in its entirety in machine-readable form.)  
\end{deluxetable}
\end{center}

\begin{center}
\begin{deluxetable}{lcccccccrrrrrrrrr}
\tabletypesize{\footnotesize}
\tablewidth{10pt}
\tablecaption{Model Atmospheric Parameters\label{tab-models}}
\tablecolumns{17}
\tablehead{
\colhead{Star} &
\colhead{\teff}&
\colhead{\teff}&
\colhead{\teff}&
\colhead{\logg}&
\colhead{\logg}&
\colhead{\teff}&
\colhead{\logg}&
\colhead{[M/H]}&
\colhead{\vmicro}&
\colhead{[Fe/H]}&
\colhead{$\sigma$}&
\colhead{\#\tablenotemark{a}}&
\colhead{[Fe/H]}&
\colhead{$\sigma$}&
\colhead{\#}&
\colhead{[Fe/H]} \\
\colhead{}                              &
\colhead{(B-V)}                           &
\colhead{(V-K)}                           &
\colhead{(LDR)}                           &
\colhead{(Hip)}                           &
\colhead{(Gaia)}                          &
\colhead{(K)}                            &
\colhead{}                            &
\colhead{}                              &
\colhead{(\kmsec)}                              &
\colhead{I}                             &
\colhead{I}                             &
\colhead{I}                             &
\colhead{II}                            &
\colhead{II}                            &
\colhead{II}                 &
\colhead{$\langle {\rm I,II}\rangle$}               
}
\startdata
HIP 258	&	4589	&	4853	&	4672	&	2.19	&	2.35	&	4551	&	2.44	&	$	0.01	$	&	1.26	&	$	0.12	$	&	0.07	&	52	&	$	0.12	$	&	0.08	&	8	&	$	0.12	$	\\
HIP 410	&	4954	&	4854	&	5022	&	2.29	&		&	4876	&	2.57	&	$	-0.18	$	&	1.15	&	$	-0.12	$	&	0.06	&	66	&	$	-0.07	$	&	0.06	&	10	&	$	-0.10	$	\\
HIP 476	&	5068	&	5329	&	5117	&	2.33	&		&	5109	&	2.68	&	$	-0.13	$	&	1.36	&	$	-0.02	$	&	0.07	&	66	&	$	0.00	$	&	0.04	&	11	&	$	-0.02	$	\\
HIP 716	&	4887	&	5010	&	4984	&	2.37	&	2.41	&	4732	&	2.11	&	$	-0.41	$	&	1.35	&	$	-0.28	$	&	0.06	&	65	&	$	-0.33	$	&	0.04	&	10	&	$	-0.31	$	\\
HIP 1009	&	4589	&	5061	&	4697	&	2.30	&	2.35	&	4631	&	2.60	&	$	0.20	$	&	1.19	&	$	0.34	$	&	0.08	&	50	&	$	0.31	$	&	0.05	&	9	&	$	0.32	$	\\
...
\enddata
\tablenotetext{a}{\#: The number of lines contributing to an abundance determination.}
\vspace{1ex}
      \item (This table is available in its entirety in machine-readable form.) 
\end{deluxetable}
\end{center}

\begin{center}                                                
\begin{deluxetable}{lccc}                         
\tabletypesize{\footnotesize}                                 
\tablewidth{0pt}                                              
\tablecaption{Sensitivity ($\sigma$) of elemental abundances
to the uncertainties in atmospheric parameters.\label{tab-uncert}}
\tablecolumns{4}                                             
\tablehead{                                                   
\colhead{Species}                          &                                         
\colhead{$\Delta$\teff (K)}                       &                     
\colhead{$\Delta$\logg}                      &                     
\colhead{$\Delta$\vmicro (\kmsec)} \\
\colhead{}                              &                                       
\colhead{$-$150/$+$150}                             &                     
\colhead{$-$0.25/$+$0.25}                             &
\colhead{$-$0.25/$+$0.25}                          
}                                                             
\startdata                                                    
\mbox{Si\,{\sc i}}	&	$+$0.01	/	$-$0.02	&	$-$0.03	/	$+$0.03	&	$+$0.04	/$-$0.04	\\
\mbox{Ca\,{\sc i}}	&	$-$0.13	/	$+$0.14	&	$+$0.03	/	$-$0.03	&	$+$0.12	/$-$0.11	\\
\mbox{Ti\,{\sc i}}	&	$-$0.18	/	$+$0.20	&	0.00	/	0.00	&	$+$0.04	/	$-$0.05	\\
\mbox{Ti\,{\sc ii}}	&	$+$0.03	/	$-$0.02	&	$-$0.11	/	$+$0.11	&	$+$0.11	/$-$0.12	\\
\mbox{Cr\,{\sc i}}	&	$-$0.16	/	$+$0.17	&	$+$0.02	/	$-$0.02	&	$+$0.09	/$-$0.08	\\
\mbox{Cr\,{\sc ii}}	&	$+$0.10	/	$-$0.08	&	$-$0.10	/	$+$0.11	&	$+$0.10	/$-$0.10	\\
\mbox{Fe\,{\sc i}}	&	$-$0.12	/	$+$0.11	&	0.00	/	0.00	&	$+$0.11	/	$-$0.11	\\
\mbox{Fe\,{\sc ii}}	&	$+$0.11	/	$-$0.11	&	$-$0.12	/	$+$0.12	&	$+$0.08	/$-$0.08	\\
\mbox{Ni\,{\sc i}}	&	$-$0.09	/	$+$0.07	&	$-$0.03	/	$+$0.03	&	$+$0.08	/$-$0.09	\\
\enddata
\end{deluxetable}
\end{center}

\begin{center}                                                
\begin{deluxetable}{lrrrrrrrrrrrrrrrrrrrrr}                         
\tabletypesize{\tiny}                                 
\tablewidth{0pt}                                              
\tablecaption{Elemental Abundances\label{tab-abunds}} 
\tablecolumns{22}                                             
\tablehead{                                                   
\colhead{Star}                          &                                         
\colhead{[Si/Fe]}                       &                     
\colhead{$\sigma$}                      &                     
\colhead{\#\tablenotemark{a}}                            &
\colhead{[Ca/Fe]}                       &                     
\colhead{$\sigma$}                      &                     
\colhead{\#}                            &
\colhead{[Ti/Fe]}                       &                     
\colhead{$\sigma$}                      &                     
\colhead{\#}                            &
\colhead{[Ti/Fe]}                       &                     
\colhead{$\sigma$}                      &                     
\colhead{\#}                            &
\colhead{[Cr/Fe]}                       &                     
\colhead{$\sigma$}                      &                     
\colhead{\#}                            &
\colhead{[Cr/Fe]}                       &                     
\colhead{$\sigma$}                      &                     
\colhead{\#}                            &
\colhead{[Ni/Fe]}                       &                     
\colhead{$\sigma$}                      &                     
\colhead{\#}                            \\
\colhead{}                              &                                       
\colhead{I}                             &                     
\colhead{I}                             &
\colhead{I}                             &
\colhead{I}                             &                     
\colhead{I}                             &
\colhead{I}                             &
\colhead{I}                             &                     
\colhead{I}                             &
\colhead{I}                             &
\colhead{II}                            &                     
\colhead{II}                            &
\colhead{II}                            &
\colhead{I}                             &                     
\colhead{I}                             &
\colhead{I}                             &
\colhead{II}                            &                     
\colhead{II}                            &
\colhead{II}                            &
\colhead{I}                             &                     
\colhead{I}                             &
\colhead{I}                                                   
}                                                             
\startdata                                                    
HIP 258	&       0.36&   0.06&   10&      0.03&   0.06&    7&
             $-$0.09&   0.07&   10&   $-$0.06&   0.09&    5&
             $-$0.02&   0.07&   13&      0.23&   0.09&    4&
                0.14&   0.07&   16\\
HIP 410	&       0.15&   0.06&   15&   $-$0.07&   0.05&   10&
             $-$0.05&   0.07&   11&   $-$0.09&   0.11&    5&
             $-$0.03&   0.07&   11&   $-$0.03&   0.05&    5&
             $-$0.02&   0.07&   23\\
HIP 476	&       0.10&   0.07&   13&   $-$0.10&   0.05&   10&
             $-$0.01&   0.08&   11&   $-$0.01&   0.04&    5&
             $-$0.04&   0.08&   15&      0.00&   0.05&    5&
             $-$0.08&   0.06&   22\\
HIP 716	&       0.17&   0.07&   16&   $-$0.13&   0.05&   10&
             $-$0.03&   0.07&   11&   $-$0.03&   0.09&    5&
             $-$0.04&   0.05&   11&   $-$0.05&   0.05&    5&
             $-$0.06&   0.06&   24\\
HIP 1009&       0.36&   0.05&   10&   $-$0.04&   0.08&    9&
             $-$0.11&   0.09&   11&   $-$0.05&   0.05&    3&
             $-$0.09&   0.07&   13&   $-$0.15&   0.07&    3&
                0.08&   0.05&   14\\
...
\enddata
\tablenotetext{a}{\#: The number of lines contributing to an abundance determination.}
\vspace{1ex}
      \item (This table is available in its entirety in machine-readable form.) 
\end{deluxetable}
\end{center}

\clearpage
\begin{center}
\begin{deluxetable}{lcccc}
\tabletypesize{\footnotesize}
\tablewidth{10pt}
\tablecaption{Galactic Orbital Parameters\label{orbpar}}
\tablecolumns{11}
\tablehead{
\colhead{Star}         &
\colhead{L$_{\rm{z}}$ (kpc $\kmsec$)}  &
\colhead{E$_{\rm{tot}}$ (km$^{2}$ s$^{-2}$)}  &
\colhead{R$_{\rm{m}}$ (kpc)}          &
\colhead{$e$}     
}
\startdata
HIP 258	&	1394.9	$^{+	18.0	}_{	-20.5	}$&	$	-88280.4	$	$^{+	359.9	}_{	-377.9	}$&	6.64	$^{+	0.10	}_{	-0.14	}$&	0.211	$^{+	0.014	}_{	-0.009	}$	\\
HIP 410	&	1636.7	$^{+	18.8	}_{	-17.1	}$&	$	-81144.0	$	$^{+	256.7	}_{	-176.4	}$&	7.62	$^{+	0.13	}_{	-0.16	}$&	0.180	$^{+	0.014	}_{	-0.013	}$	\\
HIP 476	&	1715.8	$^{+	2.7	}_{	-3.6	}$&	$	-80595.7	$	$^{+	63.3	}_{	-86.9	}$&	7.75	$^{+	0.04	}_{	-0.04	}$&	0.050	$^{+	0.003	}_{	-0.003	}$	\\
HIP 716	&	1781.0	$^{+	8.4	}_{	-11.0	}$&	$	-76473.0	$	$^{+	201.7	}_{	-148.2	}$&	8.32	$^{+	0.19	}_{	-0.17	}$&	0.207	$^{+	0.016	}_{	-0.013	}$	\\
HIP 1009	&	1315.9	$^{+	11.0	}_{	-17.0	}$&	$	-89667.2	$	$^{+	151.4	}_{	-237.5	}$&	6.44	$^{+	0.10	}_{	-0.16	}$&	0.269	$^{+	0.019	}_{	-0.010	}$	\\
...
\enddata
\vspace{1ex}
      \item (This table is available in its entirety in machine-readable form.) 
\end{deluxetable}
\end{center}

\end{document}